\documentclass[useAMS,usenatbib]{mn2e}
\usepackage{natbibmnfix}
\usepackage{graphicx}
\usepackage{grffile}
\usepackage{subfigure}
\usepackage{mathrsfs,amssymb}
\usepackage{amsmath}
\usepackage{natbib}
\usepackage{color}
\usepackage{ulem}
\usepackage{url}
\usepackage{bm}
\usepackage{multirow}
\usepackage[flushleft]{threeparttable}
\usepackage{gensymb}

\newcommand{\aj}{\rm AJ}                   
\newcommand{\araa}{\rm {ARA\&A}}             
\newcommand{\apj}{\rm {ApJ}}                 
\newcommand{\apjl}{\rm {ApJ}}                
\newcommand{\apjs}{\rm {ApJS}}               

\newcommand{\mnras}{\rm {MNRAS}}             
\newcommand{\nat}{\rm {Nature}}              

\newcommand{\physrep}{\rm {Phys.~Rep.}}   

\def\gtsima{$\; \buildrel > \over \sim \;$}
\def\ltsima{$\; \buildrel < \over \sim \;$}
\def\gsim{\lower.5ex\hbox{\gtsima}}
\def\lsim{\lower.5ex\hbox{\ltsima}}
\def\simgt{\lower.5ex\hbox{\gtsima}}
\def\simlt{\lower.5ex\hbox{\ltsima}}
\def\simpr{\lower.5ex\hbox{\prosima}}

 \newcommand*\oline[1]{%
  \vbox{%
    \hrule height 0.5pt
    \kern0.25ex
    \hbox{%
      \kern-0.1em
      \ifmmode#1\else\ensuremath{#1}\fi
      \kern-0.1em
    }
  }
}

\begin{document}

\title[Faint-end of LFs]{On the faint-end of the high-$z$ galaxy luminosity function}
\author[Yue et al.]{Bin Yue$^{1}$, Andrea Ferrara$^{1, 2}$, Yidong Xu$^{3}$  \\
$^1$Scuola Normale Superiore, Piazza dei Cavalieri 7, I-56126 Pisa, Italy\\
$^2$Kavli IPMU (WPI), Todai Institutes for Advanced Study, the University of Tokyo, 5-1-5 Kashiwanoha, Kashiwa 277-8583, Japan\\
$^3$Key Laboratory for Computational Astrophysics, National Astronomical Observatories, Chinese Academy of Sciences, Beijing, 100012, China\\
}

\maketitle

\begin{abstract}
Recent measurements of the Luminosity Function (LF) of galaxies in the Epoch of Reionization (EoR, $z\gsim6$) indicate a very steep increase of the number density of low-mass galaxies populating the LF faint-end. However, as star formation in low-mass halos can be easily depressed or even quenched by ionizing radiation, a turnover is expected at some faint UV magnitudes. Using a physically-motivated analytical model, we quantify reionization feedback effects on the LF faint-end shape. We find that if reionization feedback is neglected, the power-law Schechter parameterization characterizing the LF faint-end remains valid up to absolute UV magnitude $\sim -9$. If instead radiative feedback is strong enough that quenches star formation in halos with circular velocity smaller than 50 km s$^{-1}$, the LF starts to drop at absolute UV magnitude $\sim -15$, i.e. slightly below the detection limits of current (unlensed) surveys at $z\sim5$. The LFs may rise again at higher absolute UV magnitude, where, as a result of interplay between reionization process and galaxy formation, most of the galaxy light is from relic stars formed before the EoR. We suggest that the galaxy number counts data, particularly in lensed fields, can put strong constraints on reionization feedback. In models with stronger reionization feedback, stars in galaxies with absolute UV magnitude higher than $\sim -13$ and smaller than $\sim - 8$ are typically older. Hence, the stellar age - UV magnitude relation can be used as an alternative feedback probe. 
\end{abstract}

\begin{keywords}
cosmology: observations - galaxies: high-redshift - cosmology: theory - dark ages, reionization, first stars
\end{keywords}

\section{Introduction}

Reionization is one of the most important processes during cosmic history. It starts around $z\sim 30 - 20$\footnote{This is the typical redshift when the 3$\sigma$ fluctuations of the cosmic density field on  the molecular hydrogen cooling mass scale collapse to form halos, see \citet{2001PhR...349..125B}. The very early and rare first stars can form from fluctuations $>3\sigma$ at much higher redshifts \citep{2006MNRAS.373L..98N,2007ApJ...667...38T}. }, when the first stars begin to form (see reviews e.g. \citealt{2002Sci...295...93A,2005SSRv..116..625C,2009Natur.459...49B}), and it is completed by  $z\sim5-6$, when virtually all the intergalactic medium (IGM) is ionized \citep{2001AJ....122.2850B,2006ARA&A..44..415F,  2010ApJ...723..869O, 2010MNRAS.408.1628S}. A deeper understanding of the reionization process relies on observations of its driving sources that are generally believed to be high-$z$ normal star-forming galaxies \citep{2011MNRAS.414..847S, 2013ApJ...768...71R,2011MNRAS.414.1455L,2012MNRAS.420.1606J,2012ApJ...758...93F, 2013ApJ...768...71R,2015ApJ...802L..19R}. So far, observations of high-$z$ LFs have reached absolute UV magnitude $\sim -13$ (at $z\sim6$) and redshift up to $\sim10$ \citep{2015ApJ...803...34B,2013MNRAS.432.2696M,2014ApJ...786..108O, 2015ApJ...810...71F, 2015arXiv150601035B, 2015arXiv150600854R} thanks to the deepest HUDF/XDF surveys \citep{2013ApJS..209....6I,2011ApJ...737...90B,2013ApJ...773...75O,2013ApJ...763L...7E} and gravitational lensing magnification \citep{2015ApJ...800...18A,2016arXiv160205199M,2016arXiv160406799L}. Recent observations of high-$z$ galaxies are reviewed in \citet{2015arXiv151105558F}.

The observed high-$z$ galaxies are mostly the brightest ones among reionization sources. However, in the hierarchical structure formation scenario, massive halos form through a series of mergers of smaller progenitors. Such small halos are more numerous and dominate the collapsed matter fraction budget. Thus, it is natural to expect that more faint galaxies exist, hosting most of the stellar mass and dominating the ionizing photon budget (e.g. \citealt{2007MNRAS.380L...6C}). 

A problem might arise with the above scenario. Sustaining star formation requires continuous cold gas supply, but the available gas content  in halos could be limited by various feedback effects, namely supernova feedback and radiative feedback. Both supernova explosions and ionizing radiation inject thermal energy into the interstellar medium, interrupting the gas cooling process. However, the internal feedback due to stars is self-regulated since a decreasing star formation efficiency also reduces the feedback strength. Thus, feedback operates more efficiently when it is driven by ionizing radiation from external sources (nearby galaxies and/or a background). External ionizing photons could completely quench the star formation by either heating the cosmic gas and reducing the efficiency of gas accretion, or by continuously evaporating the gas already contained in  halos. These processes are collectively referred to as ``reionization feedback'' as they occur during the Epoch of Reionization (EoR). The effects mainly depend on the gravitational potential of individual halos, and are particularly evident in smaller galaxies. If reionization feedback is indeed effective, a significant decline in the abundance of galaxies with the lowest luminosities is expected \citep{2014MNRAS.442.2560W,2015ApJ...807L..12O}. It follows that the role played by faint galaxies as reionization sources might be questioned.

The shape of the luminosity function (LF) faint-end during EoR is expected to carry signatures of the star formation modulation imposed by impinging ionizing radiation. However, compared with the wealth of detailed theoretical studies on how external ionizing radiation suppresses star formation in galaxies (see e.g. \citealt{2013MNRAS.432.3340S,2013MNRAS.432L..51S}), there are fewer predictions on the LF of high-$z$ galaxies in the full absolute magnitude range, including the faint/low-mass  galaxies most sensitive to reionization feedback. This is clearly due to the limited dynamical range that numerical simulations can achieve. Nevertheless, \citet{2016arXiv160307729G} has numerically investigated the LF down to the faint-end. He found evidence for a deviation of the LF from the Schechter function at absolute UV magnitude $\sim -14$, although the LF continues to rise up to  magnitude $\sim-12$. This can be translated into an equivalent sharp cutoff of the Schechter function in the UV magnitude range $\sim -14$ to $\sim -12$ to match the required UV emissivity for completing the reionization \citep{2015ApJ...802L..19R, Mitra15}.

Observationally, the minimum mass of host halos of high-$z$ galaxies has been constrained by \citet{2011ApJ...729...99M}. By assuming that the galaxy stellar mass is gathered in halo growth history through a series of instantaneous star formation bursts triggered by mergers, and comparing the theoretical LFs with deep {\tt HST} surveys, they derived a minimum mass of $\approx2.5\times10^9~M_\odot$ at $z\sim6 - 8$. This minimum mass is consistent with the reionization requirements. Because of the existence of this mass floor, LFs drop gently above a turnover absolute UV magnitude. They also found that the total star formation rate (SFR) in all high-$z$ galaxies is only moderately higher than the SFR contained by already observed galaxies, therefore the star formation activity in ultra-faint galaxies must be heavily suppressed by the reionization feedback (however, for a slightly different view see \citealt{Salvadori09}).

Within current detection limits no deviation with respect to the Schechter power-law increase has been reported. However, a sensitivity improvement, as provided by, e.g., the Frontier Fields\footnote{\url{http://www.stsci.edu/hst/campaigns/frontier-fields/}},  might change the situation dramatically \citep{2014MNRAS.443L..20Y}. In view of these developments it is therefore timely to develop an analytical model of the high-$z$ galaxy LF matching the already observed bright part of the LF, and simultaneously predicting the faint-end far below the current detection limit. By comparing the predictions with upcoming observations,  we may then characterize reionization feedback effects in detail. This task involves various complications, including the metallicity and stellar evolution, and a mass-dependent star formation efficiency. Moreover, the SFR probably depends on other properties of the halo and its environments as well, in addition to the halo mass, as found by simulations (e.g. \citealt{2014MNRAS.440.2498P}.)

Pioneering works have derived the high-$z$ galaxy properties from the better-known halo properties, using various models to associate the star formation history to the mass assembly history of the host halo (e.g. \citealt{2007MNRAS.380L...6C,2010ApJ...714L.202T,2013ApJ...768L..37T,2014MNRAS.439.1326W,2015arXiv150801204M,2015ApJ...799...32B,2016MNRAS.460..417S,2016MNRAS.455.2101M}). In particular, in a series of works \citet{2010ApJ...714L.202T}, \citet{2013ApJ...768L..37T} and \citet{2015arXiv150801204M}, an analytical method has been developed to calculate the LF of high-$z$ galaxies from the dark matter halo mass function. The main assumption in their approach  is that the ``star formation efficiency" \--- defined as the ratio of gas converted into stars to  the accreted {\it dark matter} mass \--- is a function of halo mass only, i.e. it is independent of redshift. The redshift dependence of the galaxy luminosity is ascribed to its mass assembly history. Therefore, once the star formation efficiency at a given redshift is obtained, it is assumed to hold at any redshifts. Alternatively, to derive the high-$z$ galaxy LF from halo mass function, a mean redshift-independent SFR - halo mass relation is used in \citet{2016MNRAS.455.2101M}. Moreover, \citet{2015ApJ...799...32B} assumed that the specific star formation rate is proportional to the specific halo accretion rate. They found that the  stellar mass to halo mass ratio evolves rapidly at $z > 4$.
 
In this paper\footnote{Throughout the paper, we use the {\tt Planck} cosmology parameters \citep{2015arXiv150201589P}: $\Omega_m$=0.308, $\Omega_\Lambda$=0.692, $h$=0.6781, $\Omega_b$=0.0484, $n=0.9677$ and $\sigma_8$=0.8149. The transfer function is from \cite{1998ApJ...496..605E}.}, we extend the \citet{2015arXiv150801204M} model by including the reionization feedback effect under various assumptions for the feedback strength, then use it to derive the EoR galaxy LF to below current detection limits. The layout is as follows. In Section \ref{methods} we introduce the algorithm to compute LFs from halo mass functions and halo star formation histories.  We model the influence of reionization feedback on the galaxy abundance self-consistently. In Section \ref{results} we present our predicted LFs extended to galaxies in which star formation is significantly influenced by reionization feedback effects. We also discuss the SFR - stellar mass relations and the mean stellar age - UV magnitude relations there. Conclusions and discussions are given in Section \ref{conclusions}.

\section{method}\label{methods}

\subsection{From halos to galaxies}
Whether and how a halo contributes to the galaxy LF depends on how efficient it accretes matter and converts accreted matter into stars. Assume that a halo with mass $M_{\rm h}$ at redshift $z_0$ collects mass at a rate of $dM_{\rm h}/dt^\prime$ at cosmic time $t^\prime$, and suppose further that an accreted mass element, $\Delta M^\prime_{\rm h}=(dM_{\rm h}/dt^\prime)dt^\prime$, later  on increases the SFR at cosmic time $t$ by $\Delta {\rm SFR} (M_{\rm h},t,t^\prime)= f(M_{\rm h})g(t- t^\prime)\Delta M^\prime_{\rm h}$, where  the  star formation efficiency $f(M_{\rm h})$ refers to the fraction of accreted mass covered into stars. Following the spirit of \citet{2010ApJ...714L.202T}, \citet{2013ApJ...768L..37T} and \citet{2015arXiv150801204M}, we assume $f(M_{\rm h})$ depends only on halo mass $M_{\rm h}$, but we introduce an additional mass-independent function $g(t-t^\prime)$ to account for the time-dependence. Integrating over all accreted matter elements before $t$, we have the SFR of this halo at cosmic time $t$:
\begin{equation}
{\rm SFR}(M_{\rm h},t) = \int_{t_f}^{t} f(M_h)g(t- t^\prime)\frac{dM_{\rm h}}{dt^\prime}dt^\prime,
\label{SFRz}
\end{equation}
where $t_f$ is the cosmic time corresponding to the redshift $z_f$ when this halo formed. An accreted gas element is gradually converted into stars in an ``extended burst'', for which the time-dependence can be modeled as
\citep{1992ApJ...399L.113C,1996ApJ...456....1G,2000ApJ...534..507C,2002MNRAS.336L..27R,2007MNRAS.377..285S}:
\begin{equation}
g(t-t^\prime)=\frac{t- t^\prime}{\kappa^2 t^2_ d(z_f)}{\rm exp}\left[-\frac{t- t^\prime}{\kappa t_d(z_f)}\right],
\label{gdef}
\end{equation}
where 
\begin{equation}
t_d(z_f)=\sqrt{\frac{3\pi}{32G\rho_{\rm vir}(z_f)} }
\end{equation}
is the dynamic timescale for a halo with mean density $\rho_{\rm vir}$, and $\kappa$ is a free parameter that controls the duration of the burst.

Following \citet{2013ApJ...768L..37T}, we only track the halo assembly history back to the half-mass redshift, which is a good approximation as pointed by \citet{2015arXiv150801204M} (see their Fig. 4): for halos with $M_{\rm h} = 10^{11}~M_\odot$ at $5\lsim z \lsim 20 $, stars formed in progenitors with $M<M_{\rm h}/2$ only contribute $\lsim$1\% - 10\% to the total UV luminosity. A further assumption is that the halo mass grows (by accretion and minor mergers) at a constant rate between $z_0$ and $z_f$, i.e.
\begin{equation}
\frac{dM_{\rm h}}{dt^\prime} = \frac{M_{\rm h}}{ 2(t_0-t_f) },
\end{equation}
where $t_0$ is the cosmic time at redshift $z_0$. Therefore the SFR in this halo can be written as  
\begin{align}
{\rm SFR}(M_{\rm h},t) &=f(M_h)\frac{M_h}{2(t_0-t_f)}\int_{t_f}^{t} g(t- t^\prime) dt^\prime \nonumber \\
&={\rm SFR_{M15}}\int_{t_f}^{t} g(t- t^\prime) dt^\prime,
\end{align}
where SFR$_{\rm M15}$ is the SFR used in \citet{2015arXiv150801204M}. Compared with the SFR$_{\rm M15}$, the parameter $\kappa$ in the time-dependent factor $\int_{t_f}^{t} g(t- t^\prime) dt^\prime$ introduces a new degree of freedom in the SFR prescription. If $g(t-t^\prime)$ is a Dirac delta-function, which means that accreted gas forms stars instantaneously, or $\kappa \ll 1$, or $t-t_f \gg \kappa t_d$, this factor approaches unity and the SFR reduces to SFR$_{\rm M15}$. Nevertheless, $\kappa$ can be constrained only from observations (see below).
 
At redshift $z_0$, a halo of mass $M_{\rm h}$ formed at $z_f$  has a luminosity (e.g. \citealt{2007MNRAS.377..285S}):
\begin{equation}
L(M_{\rm h},z_0,z_f) = \int_{t_f}^{t_0} {\rm SFR}(M_{\rm h},t)l_\nu(t_0-t)dt,
\label{Lz}
\end{equation}
a stellar mass 
\begin{equation}
m_\star(M_{\rm h},z_0,z_f) = \int_{t_f}^{t_0} {\rm SFR}(M_{\rm h},t)m(t_0-t)dt,
\label{mstar}
\end{equation}
and a mean stellar age
\begin{equation}
t_\star(M_{\rm h},z_0,z_f)=\frac{1}{m_\star}\int_{t_f}^{t_0} {\rm SFR}(M_{\rm h},t)m(t_0-t)(t_0-t)dt.
\label{tstar}
\end{equation}
Here $l_\nu(\Delta t)$ is the Single Stellar Populations (SSP) SED template, i.e. the luminosity of an instantaneous star formation burst with unit stellar mass at the time $\Delta t$ after the burst; we always use the luminosity at the wavelength 1600~\AA~when calculating the absolute UV magnitude.  $m(\Delta t)$ is the mass fraction of surviving stars $\Delta t$ after the burst. For both $l_\nu(\Delta t)$ and $m(\Delta t)$ we use the template of \citet{2003MNRAS.344.1000B} with Chabrier IMF between 0.1 - 100~$M_\odot$ and fixed metallicity of $0.02~Z_\odot$.

Before deriving the LF of galaxies from halo mass function, we need to obtain the $f(M_h)$ curve. This is determined by comparing the theoretical mean luminosity - halo mass relations with that from observations. From Eq. (\ref{Lz}), halos with the same mass could have different luminosities if they formed at different redshifts. Given the probability distribution of the formation time, $p(w)$, their mean luminosity is 
\begin{equation}
\bar {L}(M_{\rm h},z_0) = \int L(M_{\rm h},z_0,z_f)p(w) dw,
\label{Lmean}
\end{equation}
where $w$ is related to the formation redshift $z_f$ by
\begin{equation}
w=\sqrt{0.707}\frac{\delta_c(z_f)-\delta_c(z_0)}{\sqrt{\sigma^2(M_h/2)-\sigma^2(M_h)}},
\end{equation}
with $\delta_c(z_f)$ being the critical density contrast for spherical collapse at $z_f$ linearly extrapolated  to present time, and  $\sigma^2(M_h)$ being the variance of density fluctuations smoothed on mass scale $M_h$. In the elliptical collapse scenario the $p(w)$ has the simple expression \citep{2007MNRAS.376..977G}:
\begin{equation}
p(w)=2w\,{\rm erfc}(w/\sqrt{2}),
\label{pw}
\end{equation}
where erfc is the complementary error function.

On the other hand, a relation between the observed absolute UV magnitude and the halo mass could be constructed by ``abundance matching", i.e. forcing the number density of galaxies with absolute UV magnitude smaller than $M_{\rm UV,obs}$ to match the number density of halos with mass above $M_{\rm h}$:
\begin{equation}
\int^{M_{\rm UV,obs}}_{-\infty}  \Phi_{\rm Sch}( M^\prime_{\rm UV},z_0)  dM^\prime_{\rm UV}   = \int_{M_{\rm h}}^\infty \frac{dn_{\rm h}}{dM_{\rm h}^\prime}dM_{\rm h}^\prime,
\label{match}
\end{equation}
where $\Phi_{\rm Sch}$ is the Schechter parameterization of the LF \citep{2015ApJ...803...34B}. Here the observed absolute UV magnitude $M_{\rm UV,obs}$ corresponds to the dust-attenuated luminosity, which is related to the intrinsic UV magnitude $M_{\rm UV}$ by
\begin{equation}
M_{\rm UV}=M_{\rm UV,obs} - A_{1600},
\end{equation}
where $A_{1600}$ is the dust extinction given by \citep{1999ApJ...521...64M}
\begin{equation}
A_{1600}=4.43+1.99\beta,
\end{equation} 
with $\beta$ being the luminosity-dependent UV spectrum slope.
$A_{1600}$ must be $\ge 0$ and $\beta$ is fitted in \citet{2014ApJ...793..115B} by a linear form
\begin{equation}
\beta=\beta_0+\frac{d\beta}{dM_0}(M_{\rm UV,obs}-M_0),
\end{equation}
where $M_0 = -19.5$. The best-fitted parameters of a collection of {\tt HST} observations given by \citet{2014ApJ...793..115B} are used in this work, i.e. $\beta_0 = (-1.70,-1.85, -1.91, -2.00, -2.05, -2.13)$ and $d\beta/dM_0 = (-0.20,-0.11, -0.14, -0.20, -0.20, -0.15)$ at redshifts $z=(2.5, 3.8, 5.0, 5.9, 7.0, 8.0)$, respectively. At intermediate redshifts we linearly interpolate; for higher redshifts we use the linear extrapolation for $\beta_0$ and fix $d\beta/dM_0 = -0.20$. At each $M_h$, by equating $\bar {L}$ with the luminosity of the dust-corrected absolute UV magnitude $M_{\rm UV}$, we obtain the $f(M_{\rm h})$.

After calibration, we convert the luminosity of the halo into an observed absolute UV magnitude, for which we write its explicit form here for convenience:
\begin{align}
&M_{\rm UV,obs} (M_h,z_0,z_f) = \nonumber \\
&\begin{cases}
\left( 1-1.99\frac{d\beta}{dM_0} \right)^{-1}\left(M_{\rm UV}-M_0+4.43+1.99\beta_0\right)+M_0 ,\\
~~~~~~~~({\rm if}~M_{\rm UV} < M_0- \left(  1.99\frac{d\beta}{dM_0}\right)^{-1}\left(  4.43+1.99\beta_0 \right) )\\
M_{\rm UV}, \\
~~~~~~~~({\rm if}~M_{\rm UV} \ge M_0- \left(  1.99\frac{d\beta}{dM_0}\right)^{-1}\left(  4.43+1.99\beta_0 \right) ), 
\end{cases}
\label{MUVobs}
\end{align}
where $M_{\rm UV}$ is the absolute UV magnitude of the luminosity given by Eq. (\ref{Lz}).

Galaxies with the same luminosity could be hosted by halos with different masses that formed at different redshifts. Among halos of mass $M_h$, only those formed at a specified redshift, $z_{\rm mag}$, can host galaxies with observed absolute UV magnitude $M_{\rm UV,obs}$;  $z_{\rm mag}$ is then obtained by substituting $M_{\rm UV,obs}$, $M_h$ and $z_0$ into Eq. (\ref{MUVobs}). The LF is then written as 
\begin{equation}
\Phi(M_{\rm UV,obs},z_0) =  \int dM_{\rm h}\frac{dn_{\rm h}}{dM_{\rm h}}p(w_{\rm mag})\frac{dw_{\rm mag}}{dM_{\rm UV,obs}} ,
\label{LF}
\end{equation}
where $dn_{\rm h}/dM_{\rm h}$ is the halo mass function \citep{1999MNRAS.308..119S,2001MNRAS.323....1S}, and $w_{\rm mag}=w(M_h,z_0,z_{\rm mag})$. The integration is performed over the range of $M_h$ for which a $z_{\rm mag}$ solution exists.
 
The free parameter $\kappa$ is determined by comparing the derived LF with the observations from \citet{2015ApJ...803...34B}. We use a reduced chi-square to measure the deviation of the predicted $\Phi$ from the observations, which is defined as 
\begin{equation}
\chi^2_{\rm red}(\kappa)=\frac{1}{n-1}\sum_i \frac{ (\Phi -\Phi_i)^2}{\sigma^2_i},
\end{equation}
where $\Phi$ is the predicted LFs from Eq. (\ref{LF}) by using the best-fitted Schechter parameterization of \citet{2015ApJ...803...34B} to calibrate the $f(M_{\rm h})$ (see Eq. \ref{match}), $\Phi_i$ and $\sigma_i$ are observational points of LFs and errors in \citet{2015ApJ...803...34B}. The sum is performed on all observational points at redshifts 5, 6, 7 and 8 except for those with upper limits only; $n$ is the number of all points used. 

\begin{figure}
\centering{
\includegraphics[scale=0.4]{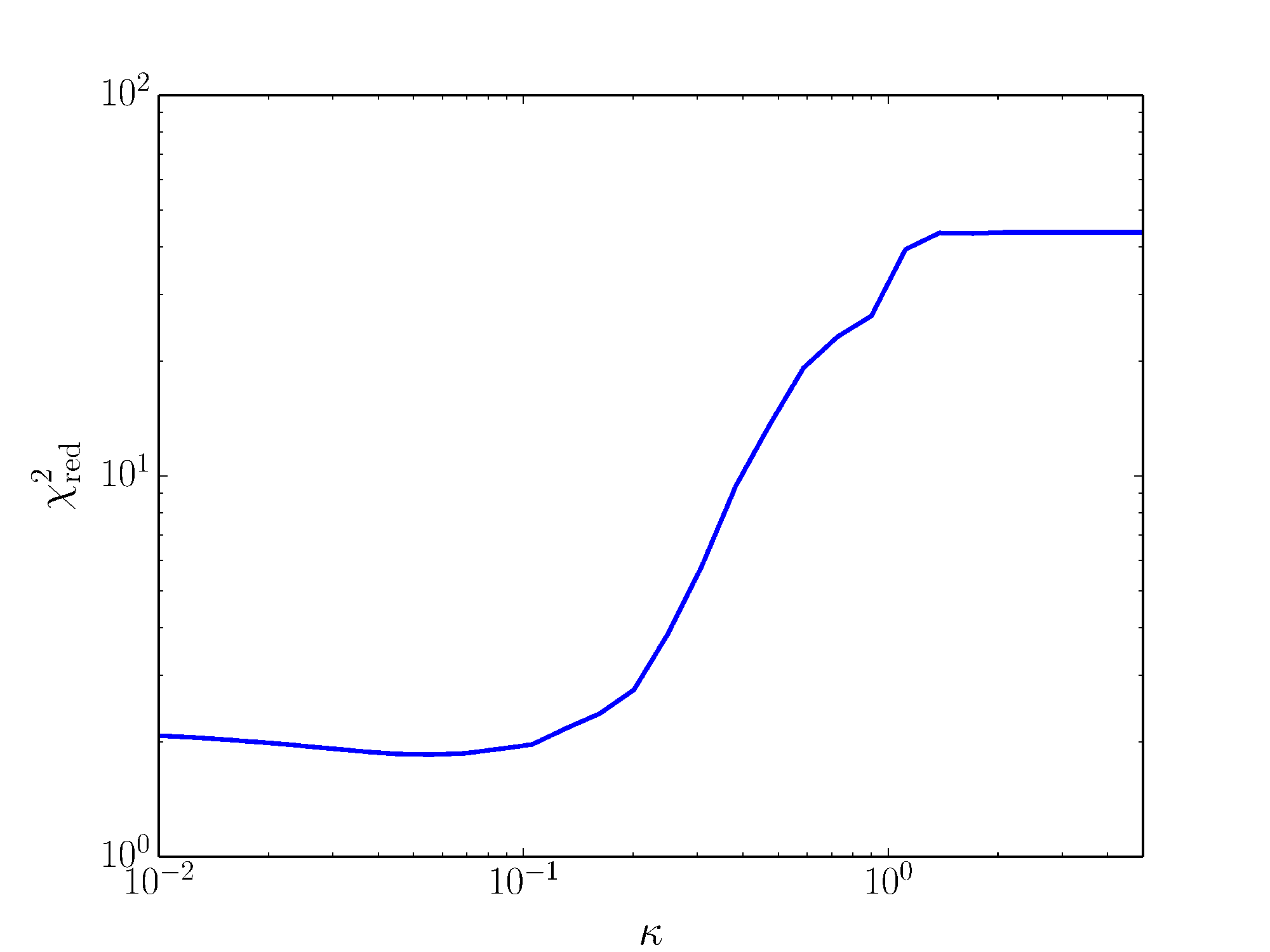}
\caption{The $\chi^2_{\rm red}$ as a function of $\kappa$.}
\label{fig_chi2_red}
}
\end{figure}

\begin{figure}
\centering{
\includegraphics[scale=0.4]{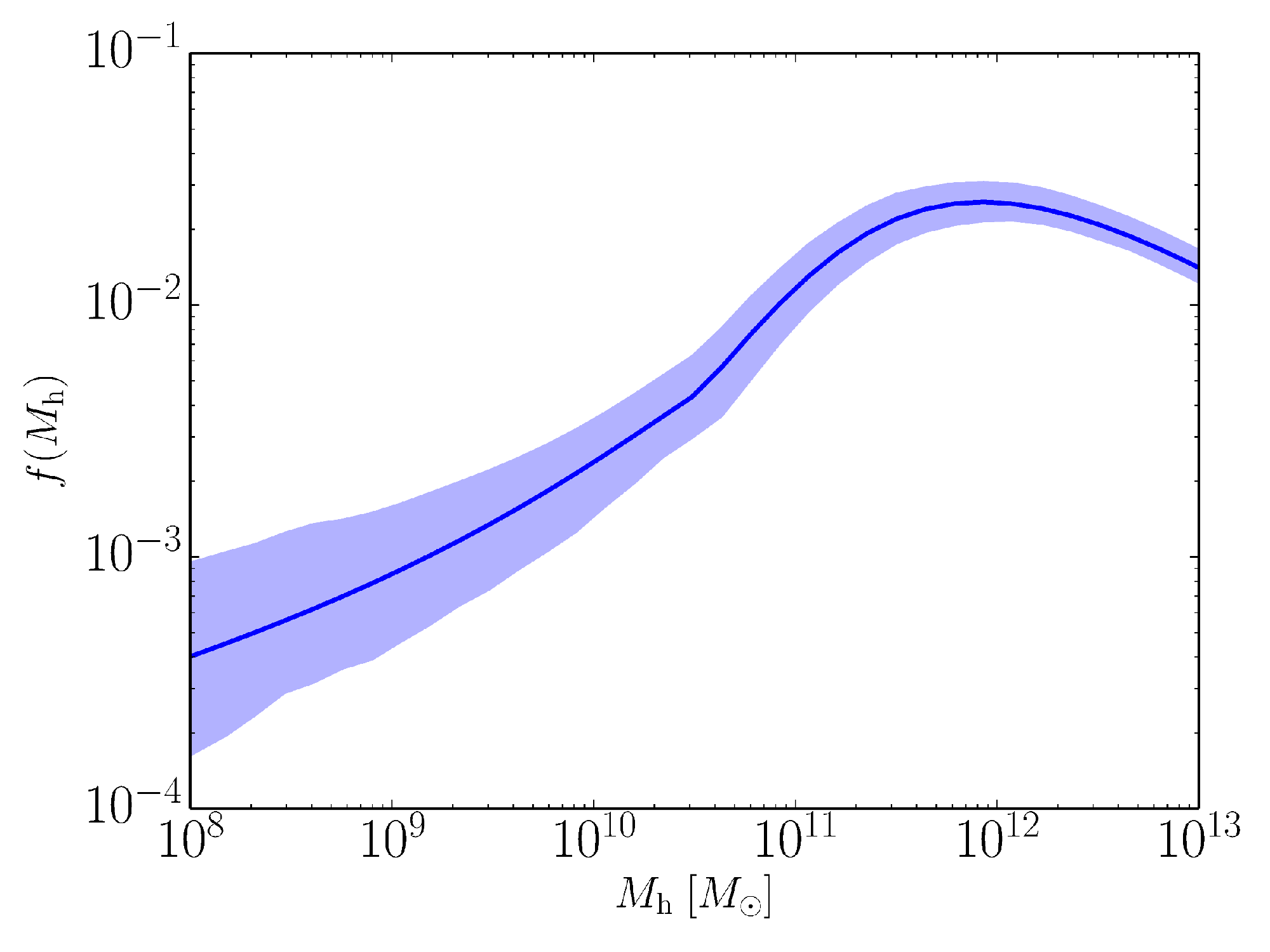}
\caption{The function $f(M_{\rm h})$ calibrated by using the Schechter formula of the observed LF at  $z_0 \sim 5$  in \citet{2015ApJ...803...34B} and fix $\kappa = 0.1$. The shaded regions are uncertainties, see the text.
}
\label{fig_fsfg}
}
\end{figure}

The $\chi^2_{\rm red}$ as a function of $\kappa$ is shown in Fig. \ref{fig_chi2_red}. We find that the deviation is small and stable at $\kappa \lsim 0.1$, implying that observations favor a scenario in which the accreted gas is typically converted into stars on a timescale  $\lsim 0.1t_d$. We do not find lower limits on the value of $\kappa$. Therefore, we take $\kappa = 0.1$ hereafter. The calibrated $f(M_{\rm h})$ at $z_0 \sim5$ is plotted in Fig. \ref{fig_fsfg}.

To derive the uncertainties in $f(M_{\rm h})$,  we randomly sample the three Schechter parameters, $\Phi^*,M_{\rm UV}^*$ and $\alpha$, assuming that they all follow a Gaussian distribution with standard deviation equal to the 1$\sigma$ error given in  \citet{2015ApJ...803...34B}. The region containing 68.3\% of the $f(M_{\rm h})$ values in each mass bin is plotted in Fig. \ref{fig_fsfg}. We do not account for the uncertainties in $\beta_0$ and $d\beta/dM_0$.  We find that the uncertainties are larger for smaller halos ($\lsim 10^{10}~M_\odot$), because of the weaker constraints on the faint-end slope of LFs. In the following, we will use the calibration at $z_0 \sim 5$ for all calculations.

\subsection{Radiative feedback during EoR}
We now incorporate the reionization feedback effects into the above modeling of galaxy LF. How reionization feedback affects a halo depends on two separate issues: (a) the probability that the halo is located in ionized bubbles,  and (b) how fast the gas supply is interrupted, or its gas photo-evaporated. For a given cosmological evolutionary scenario, (a) is determined by the total amount of emitted ionizing photons and the gas recombination rate. The point (b) is instead determined by complex radiation hydrodynamical processes inside galaxies.

Specifically, we assume that the star formation efficiency $f(M_{\rm h})$ calibrated by matching the theoretical LF to the Schechter function is valid down to the atomic-cooling halo mass (see Appendix \ref{SN} for a check of the supernova feedback effects), unless in these halos star formation (i) never ignites, or (ii) is totally quenched by the external ionizing radiation. Such events can occur in halos with circular velocity $v_c$ smaller than a threshold $v_c^*$, further located in ionized bubbles  (see e.g. \citealt{2002MNRAS.336L..27R}). Here $v_c$ depends on both $M_{\rm h}$ and $z_f$ \citep{2001PhR...349..125B}. We note that currently the threshold for quenching star formation by external ionizing radiation is still rather uncertain, and in terms of circular velocity, it varies from $\sim10$ km s$^{-1}$ to $\sim70$ km s$^{-1}$, depending on the intensity and spectrum of the ionizing flux, the time interval in which the halo is exposed to the radiation, self-shielding effects, and so on (e.g. \citealt{1996ApJ...465..608T,2004ApJ...601..666D,2000MNRAS.315L...1K,2000ApJ...542..535G,2013MNRAS.432L..51S}). 

With star formation being quenched at $z_q$, analogously to Eq. (\ref{Lz}), the emission rate of ionizing photons is
\begin{equation}
\dot{N}_{\rm ion}(M_{\rm h},z_0,z_q,z_f)= \int_{t_f}^{t_q}{\rm SFR}(M_{\rm h},t) \dot{q}(t_0-t)dt,
\label{Nion}
\end{equation}
where the rate $\dot{q}$ is taken from \citet{2003MNRAS.344.1000B}. Taking into account the reionization feedback, we have three possible star formation histories in halos with viral temperature $>10^4$~K (we neglect the contribution from minihalos, as substantial star formation activity in minihalos is unlikely and rather uncertain): 
\begin{enumerate}
\item A halo with $v_c < v_c^*$ can sustain continuous star formation if it is always located outside ionized bubbles. 
\item A halo with  $v_c < v_c^*$ that originally formed in a neutral region later became ionized has its star formation quenched\footnote{We warn that our treatment here is simplified, as $v_c^*$ might be a function of redshift  and intensity of the UV radiation field. We neglect this effect here but refer the interested readers to the discussion in \cite{2013MNRAS.432L..51S}.} at $z_q$ ($z_0 < z_q < z_f$). In this case the star formation activity only happens between $z_f$ and $z_q$.
\item A halo with $v_c>v_c^*$ is massive enough that reionization feedback has no effects on it, no matter whether it is located inside or outside ionized bubbles.
\end{enumerate}

To evaluate reionization feedback we need to know the probability, ${\mathcal P}_{b}(M_{\rm h}, z)$, that a halo with mass $M_{\rm h}$ is located in an ionized bubble at redshift $z$. This probability is closely related to the volume filling factor of ionized regions $Q_{\rm HII}$ at the corresponding redshift. Once the escape fraction of ionizing photons, $f_{\rm esc}$, is assigned, we can compute the  evolution of the filling factor of ionized regions
\begin{equation}
\frac{dQ_{\rm HII}}{dt}=f_{\rm esc}\frac{\dot{n}_{\rm ion}}{n_{\rm H}}-Q^2_{\rm HII}C(z)n_{\rm H}(1+z)^3\alpha_{\rm B},
\label{dQHII}
\end{equation}
where $\dot{n}_{\rm ion}$ is the ionizing photon emissivity, $n_{\rm H}$ is the comoving total hydrogen number density, $C(z)$ is the clumping factor, and $\alpha_{\rm B}$ is the Case B recombination coefficient. For simplicity we take $T=10^4$~K for ionized regions, yielding $\alpha_{\rm B}=2.6\times10^{-13}$~cm$^3$s$^{-1}$. We use the form for the clumping factor given by \cite{2007MNRAS.376..534I} normalized to $C=3.0$ at $z = 5$: 
\begin{equation}
C(z)=6.8345\times{\rm exp}(-0.1822z+0.003505z^2).
\end{equation}
Taking into account the three cases of star formation history described above, the ionizing photon emissivity can be written as
\begin{align}
\dot{n}_{\rm ion} (z_0)=\int dM_{\rm h} \frac{dn_{\rm h}}{dM_{\rm h}} \int dw\,p(w) \int_{z_f}^{z_0}    \dot{N}_{\rm ion}{\mathcal F}(M_{\rm h},z_q,z_f)   dz_q, 
\label{nion}
\end{align}
where ${\mathcal F}(M_{\rm h},z_q,z_f)$ is given by 
\begin{align}
{\mathcal F}=
\begin{cases}
[1-{\mathcal P}_b(M_{\rm h},z_0)]\delta(z_q-z_0) +\frac{d {\mathcal P}_b}{dz_q} & v_c < v_c^*  \\
\delta(z_q-z_0) & v_c \ge v_c^*
\end{cases}
\label{FM}
\end{align}
in which $\delta$ is the Dirac delta-function. The term $[1-{\mathcal P}_b(M_{\rm h},z_0)]\delta(z_q-z_0)$ corresponds to the case (i), and the terms $\frac{d {\mathcal P}_b}{dz_q}$ and $\delta(z_q-z_0)$ identify case (ii) and (iii), respectively. To construct a mapping from $Q_{\rm HII}$ to ${\mathcal P}_{\rm b}$ we use a method based on the excursion-set formalism (see Appendix \ref{PBvsQ}). Once $f_{\rm esc}$ and $v_c^*$ are specified, we numerically solve the differential equation (\ref{dQHII}) from $z = 20$, when it is safe enough to assume $Q_{\rm HII} \approx 0$ and ${\mathcal P}_{\rm B} \approx 0$. 

Once the ionizing history is determined, the IGM electron scattering optical depth to CMB photons can be derived from
\begin{equation}
\tau = \sigma_{\rm T}n_{\rm H} c\left(1+\frac{Y_{\rm He}}{4Y_{\rm H}}\right) \int_0^z Q_{\rm HII}(z^\prime) (1+z^\prime)^3\vert \frac{dt}{dz^\prime}\vert dz^\prime,
\end{equation}
where $\sigma_{\rm T} = 6.65\times10^{-25}$ cm$^2$ is the cross-section for Thomson scattering, $c$ is the speed of light, $Y_{\rm He} = 0.24$ is the He mass fraction; the H fraction is $Y_{\rm H} = 1- Y_{\rm He}$;  the singly ionized He fraction is assumed to be equal to the HII fraction. 

\begin{figure}
\centering{
\includegraphics[scale=0.4]{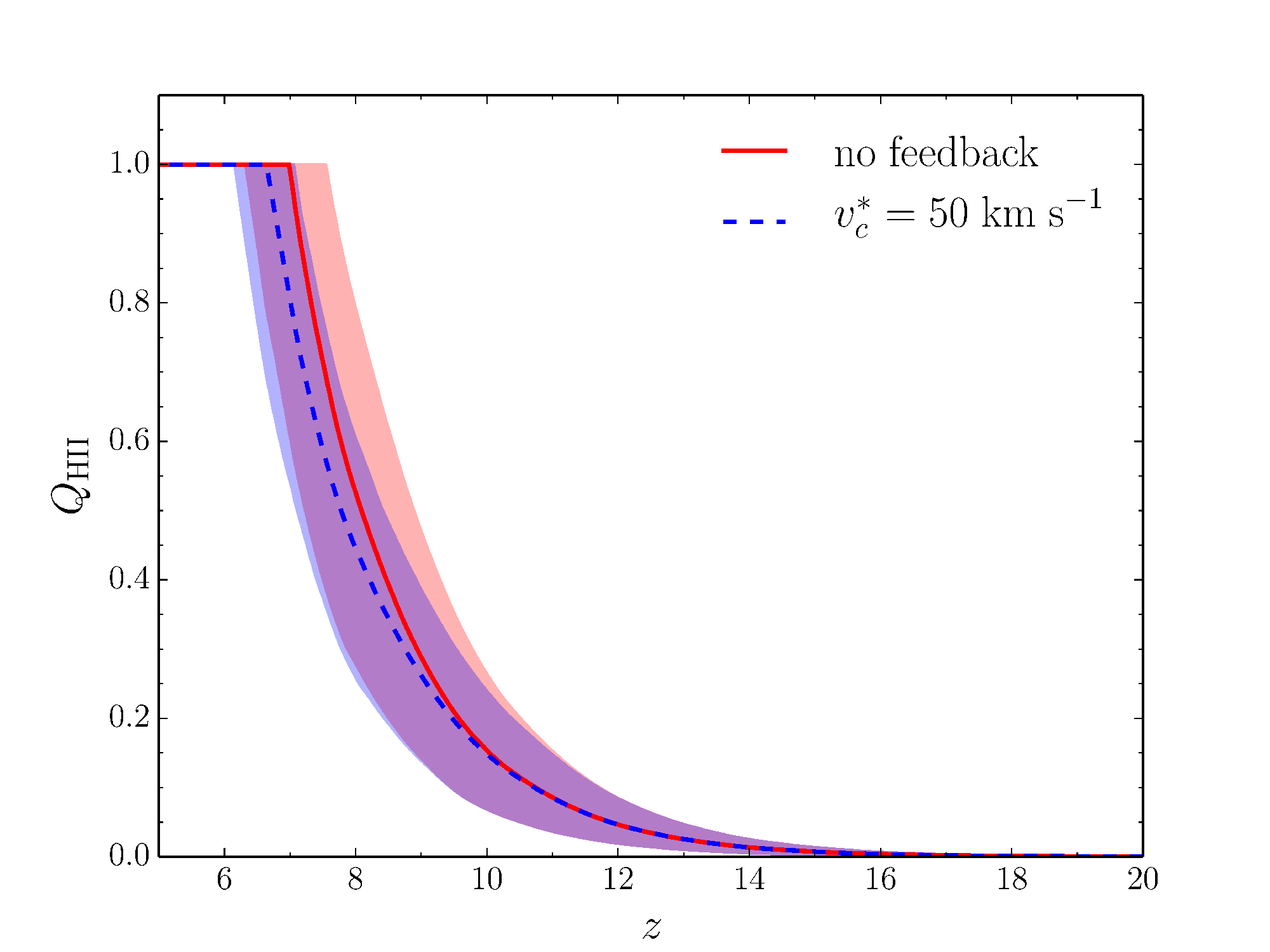}
\caption{The evolution of the filling factor of ionized regions for models without feedback (solid line) and with $v_c^* = 50$ km s$^{-1}$ (dashed line), with the shaded regions being the 68.3\% confidence interval. The corresponding e.s. optical depth values are $\tau=0.063^{+0.009}_{-0.008}$ and $\tau=0.061^{+0.008}_{-0.007}$ respectively.}
\label{fig_QHII}
}
\end{figure}

To gain a physical intuition of the effects of feedback on reionization history, in Fig. \ref{fig_QHII} we compare $Q_{\rm HII}(z)$ obtained for a no-feedback model and a strong feedback ($v_c^* = 50$ km s$^{-1}$) model, using a fixed $f_{\rm esc} = 0.2$. We find that the difference between these two models is modest. Adopting a $v_c^* = 50$ km s$^{-1}$ only reduces the Thomson optical depth by $\Delta \tau\approx0.002$ and delays the reionization by $\Delta z\approx 0.3$.  

After including the reionization feedback, the luminosity of a galaxy in which star formation is quenched at $z_q$ ($z_0 < z_q < z_f$) is then
\begin{equation}
L(M_{\rm h},z_0,z_q,z_f) = \int_{t_f}^{t_q} {\rm SFR}(M_{\rm h},t)l_\nu(t_0-t)dt,
\label{Lzq}
\end{equation}
where $t_q$ is the quenching time corresponding to redshift $z_q$. This luminosity can be converted into observed absolute UV magnitude by Eq. (\ref{MUVobs}). The LF including feedback effects is
\begin{align}
&\Phi(M_{\rm UV,obs},z_0) =\\
&  \int dM_{\rm h}\frac{dn_{\rm h}}{dM_{\rm h}}   [1-{\mathcal P}_b(M_{\rm h},z_0)]    p(w_{\rm mag})    I(v_c)    \frac{dw_{\rm mag}}{dM_{\rm UV,obs}}  \nonumber \\
&+\int dM_{\rm h}\frac{dn_{\rm h}}{dM_{\rm h}}  \int_{z_f}^{z_0} dz_q\frac{d{\mathcal P}_b}{dz_q}p(w_{{\rm mag},q})I(v_c)\frac{dw_{{\rm mag},q}}{dM_{\rm UV,obs}} \nonumber \\
&+ \int dM_{\rm h}\frac{dn_{\rm h}}{dM_{\rm h}}  p(w_{\rm mag})[1-I(v_c)]\frac{dw_{\rm mag}}{dM_{\rm UV,obs}}, 
\end{align}
where, in analogy with the no-feedback case, $w_{{\rm mag},q}$ is obtained by substituting $M_h$, $z_0$, $M_{\rm UV,obs}$ and $z_q$ into  Eq. (\ref{MUVobs}). $I$ is a step function: $I (v_c)= 1$ when $v_c < v_c^*$, and $I(v_c) = 0$ otherwise. The three terms in the above equation correspond to the three cases of star formation history. If ${\mathcal P}_b \equiv 0$  the reionization feedback is not accounted for and the above equation reduces to Eq. (\ref{LF}).

\section{results}\label{results}

\subsection{Reionization history}

We first check whether the Thomson optical depth $\tau$ is consistent with observations in this new scenario. In principle, a higher $f_{\rm esc}$ would promote the reionization process and result in a larger $\tau$. However, it also increases the probability for halos to be located in ionized bubbles, thus reducing the contribution of ionizing photons from small halos. 

\begin{figure*}
\centering{
\subfigure{\includegraphics[scale=0.28]{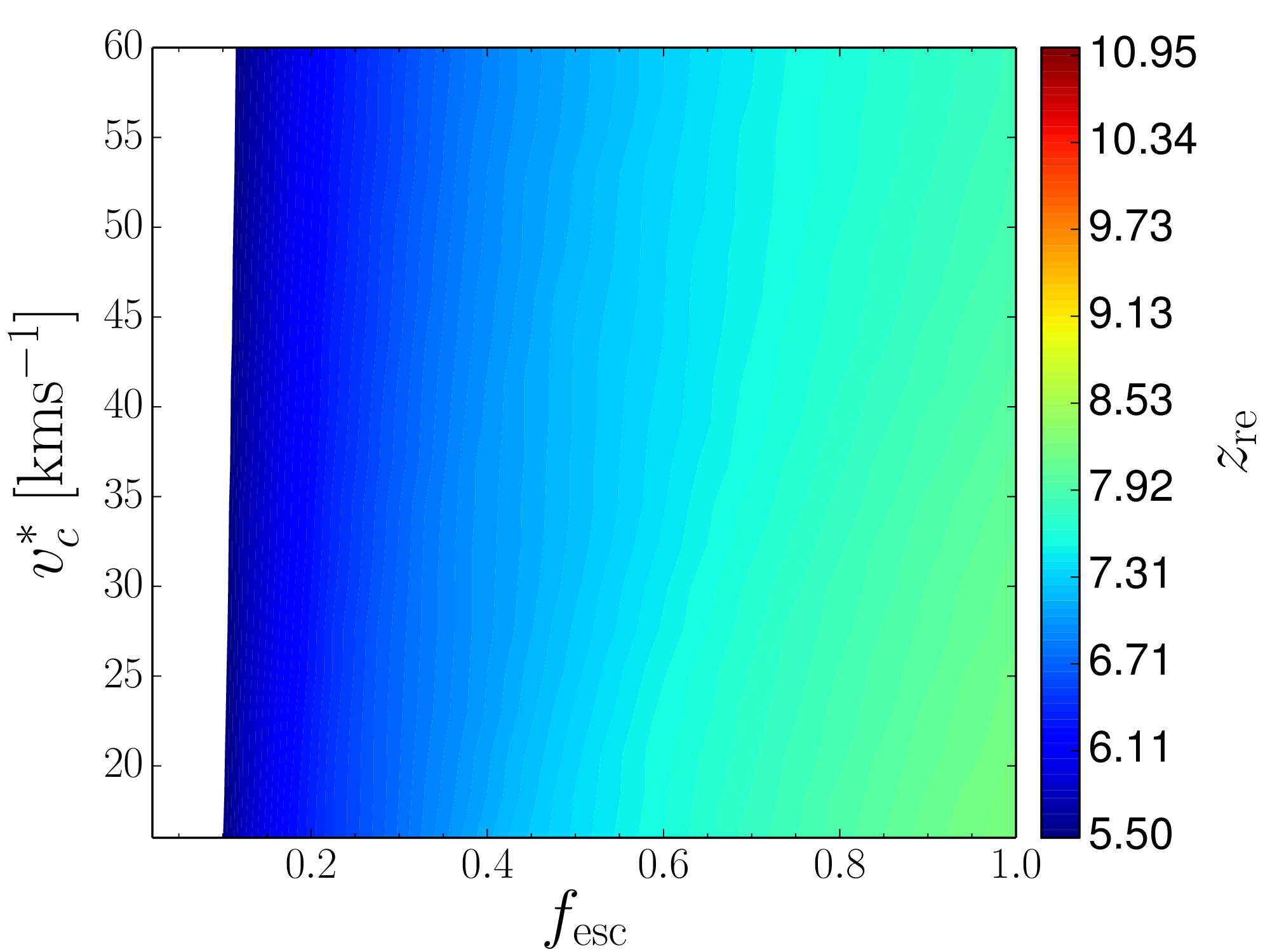}}
\subfigure{\includegraphics[scale=0.28]{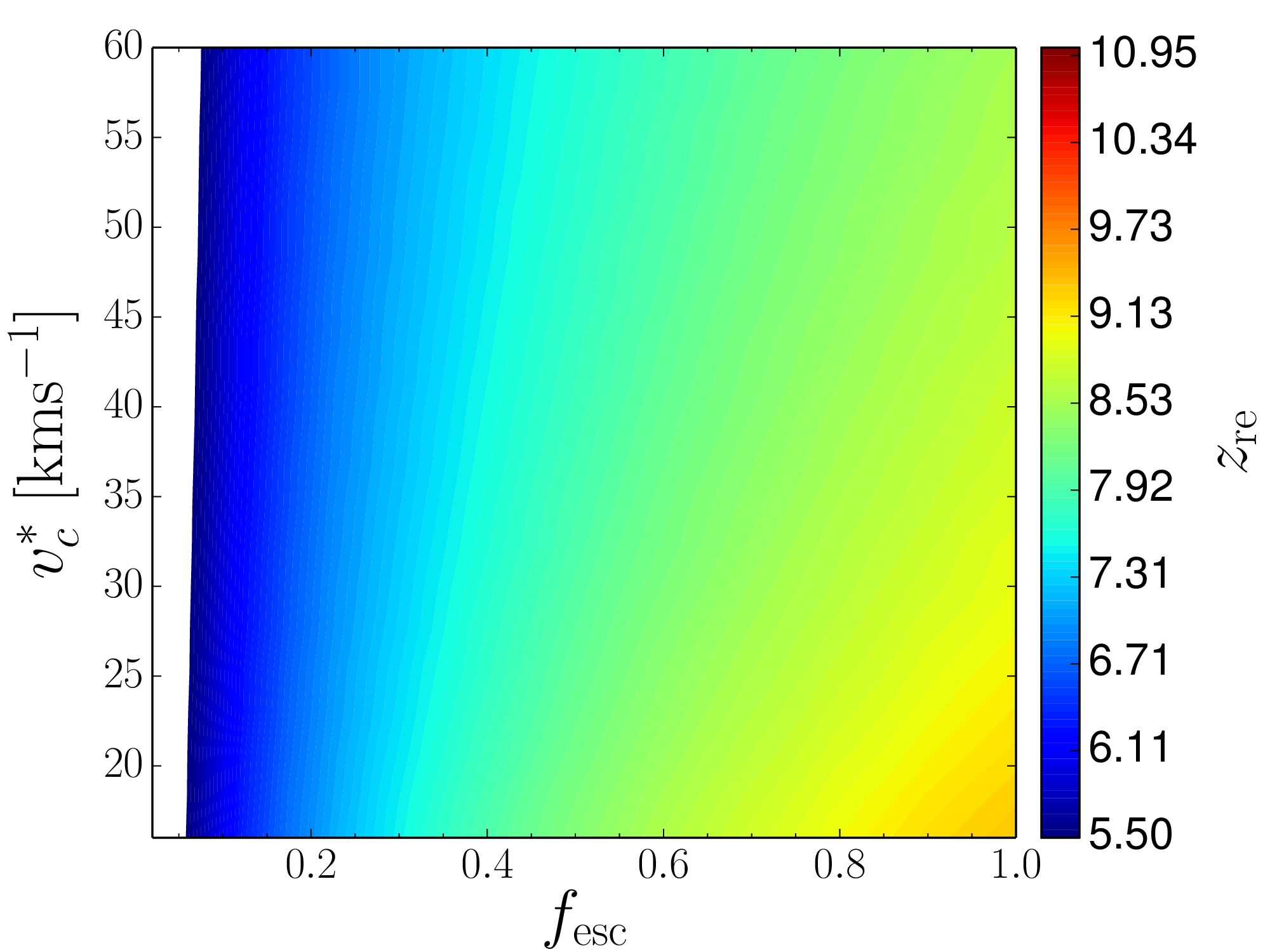}}
\subfigure{\includegraphics[scale=0.28]{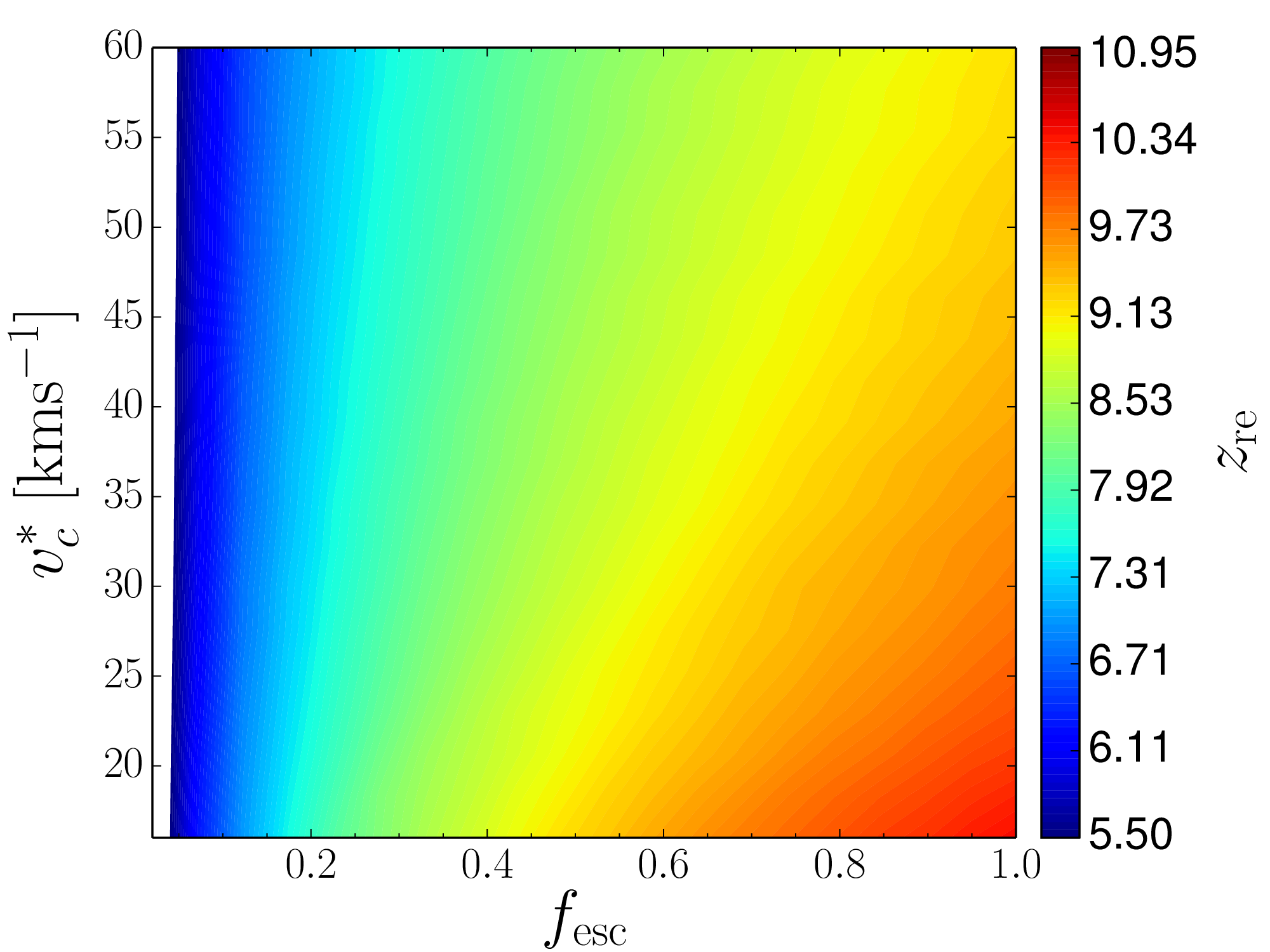}}
\subfigure{\includegraphics[scale=0.28]{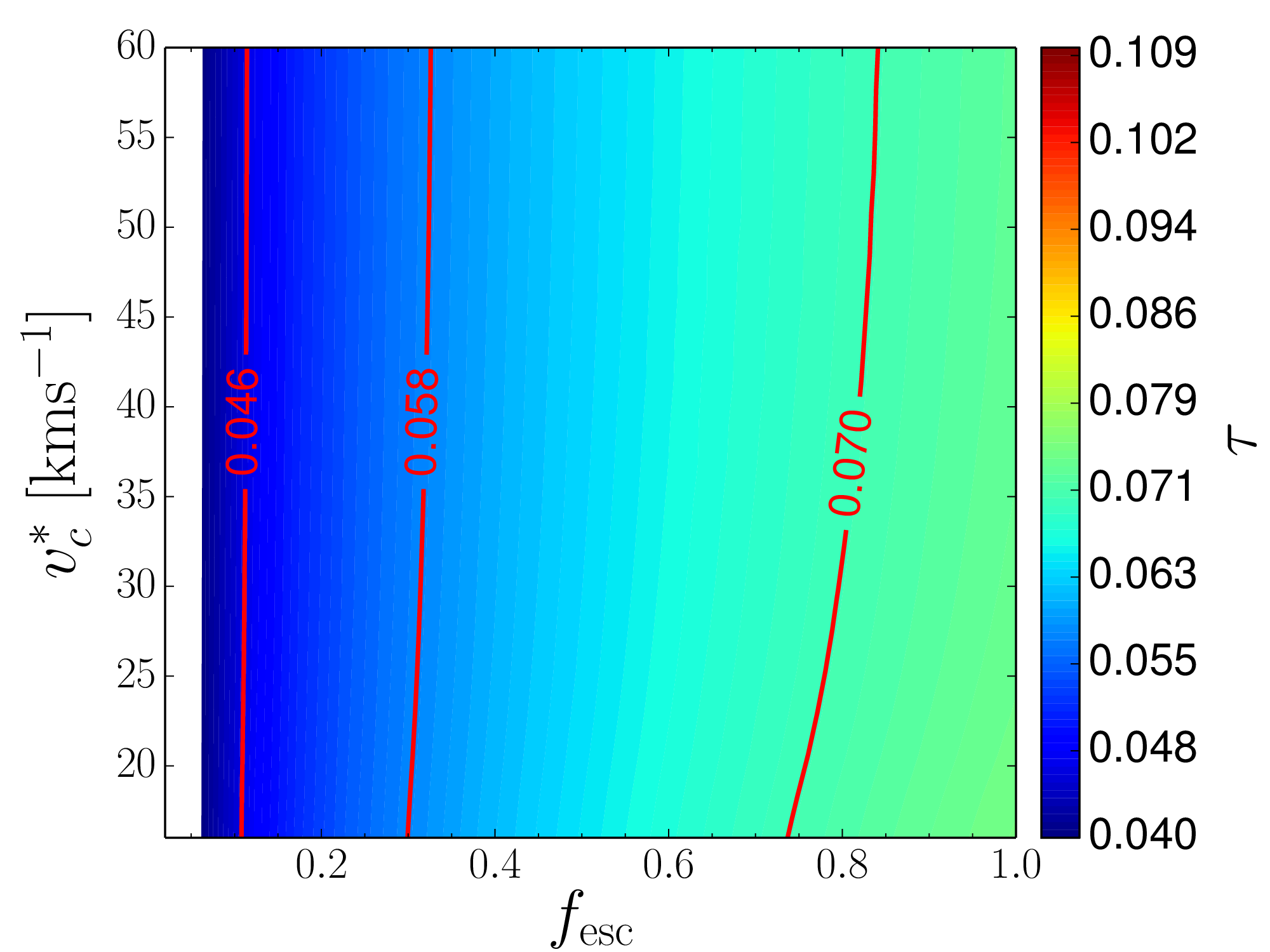}}
\subfigure{\includegraphics[scale=0.28]{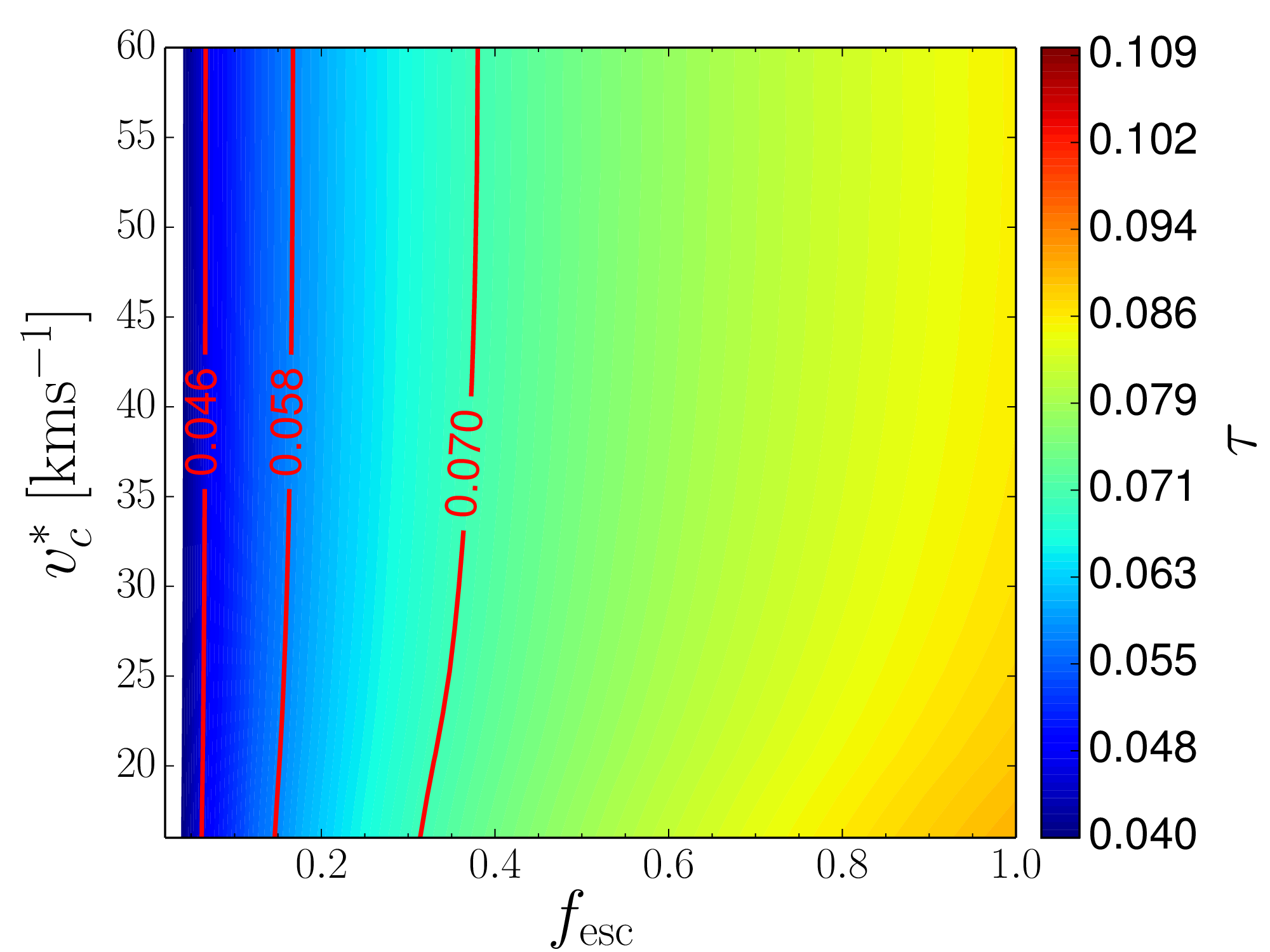}}
\subfigure{\includegraphics[scale=0.28]{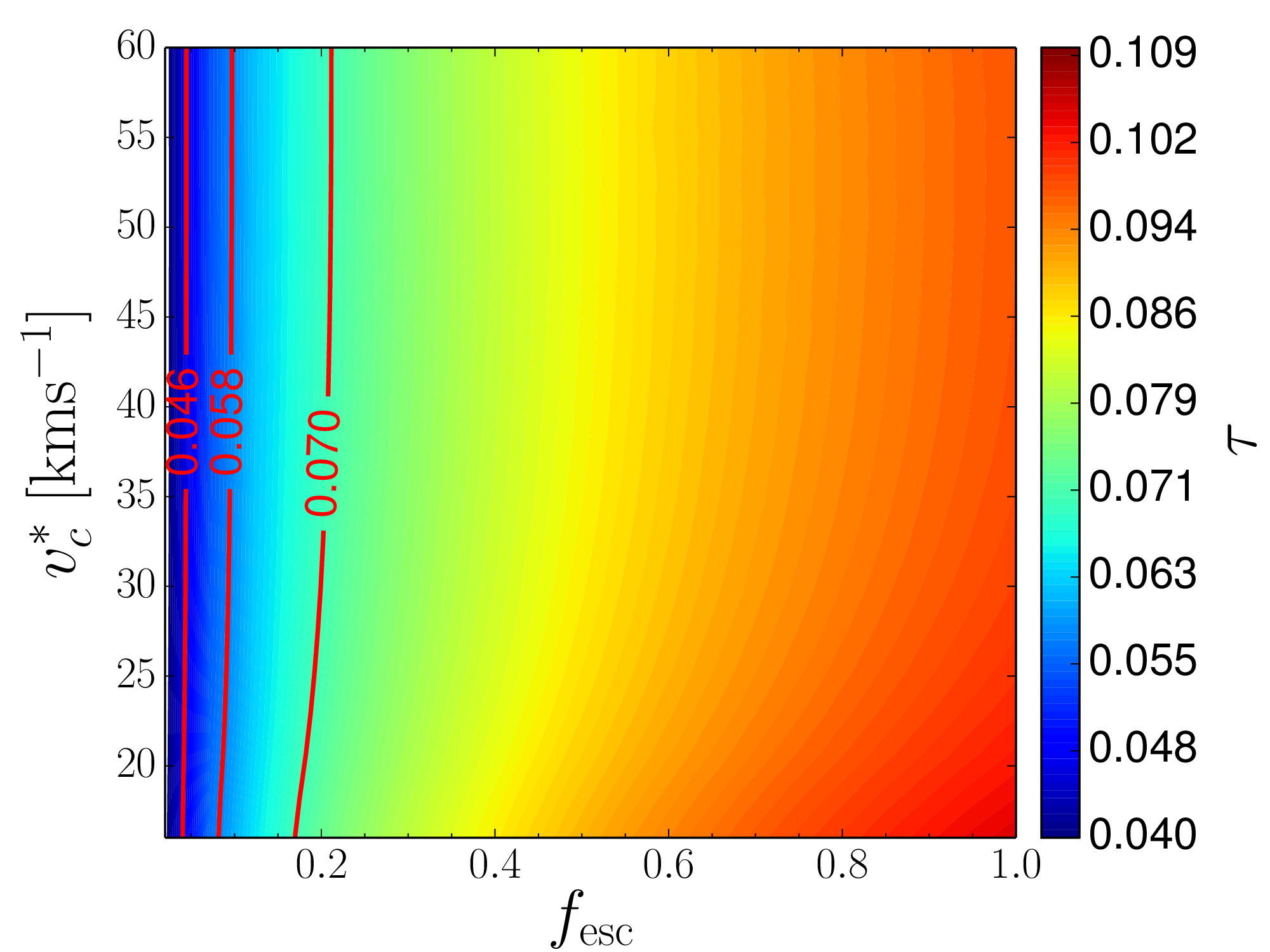}}
\caption{The completion redshift of reionization (top row) and the Thomson optical depth (bottom row) as a function of $f_{\rm esc}$ and $v_c^*$. In each row, the left and right panels are the lower and upper limits of the 68.3\% confidence intervals; in the middle panel we use best-fit Schechter parameters in the $f(M_{\rm h})$ calibration. In the bottom panels we mark the {\tt Planck} measurements  $\tau = 0.058\pm 0.012$ \citep{2016arXiv160503507P} by lines.}
\label{fig_zre_and_tau}
}
\end{figure*}

In Fig. \ref{fig_zre_and_tau} we show the completion redshift of reionization, $z_{\rm re}$, and the $\tau$ for different $f_{\rm esc}$ and $v_c^*$ pairs.  In each row, the left, central, and right panels show the minimal, best-fit, and the maximal values allowed by the uncertainty in the star formation efficiency calibration, respectively.

The Thomson optical depth $\tau$ is mainly dependent on $f_{\rm esc}$, but insensitive to  $v_c^*$. This is because not all halos with $v_c < v_c^*$ are sterile, and only those that inhabit ionized bubbles lose the ability to produce ionizing photons. In addition, star formation efficiency in small halos is low, as seen from Fig. \ref{fig_fsfg}. We find that, taking into account the uncertainties in star formation efficiency calibration, the recent {\tt Plank} measurements ($\tau=0.058\pm0.012$, \citealt{2016arXiv160503507P}) rule out both the low escape fraction range $f_{\rm esc} \lsim 0.04$ (see the right panel for $\tau$), and the high escape fraction range $f_{\rm esc}\gsim 0.8$ (left panel).

\begin{figure}
\centering{
\includegraphics[scale=0.4]{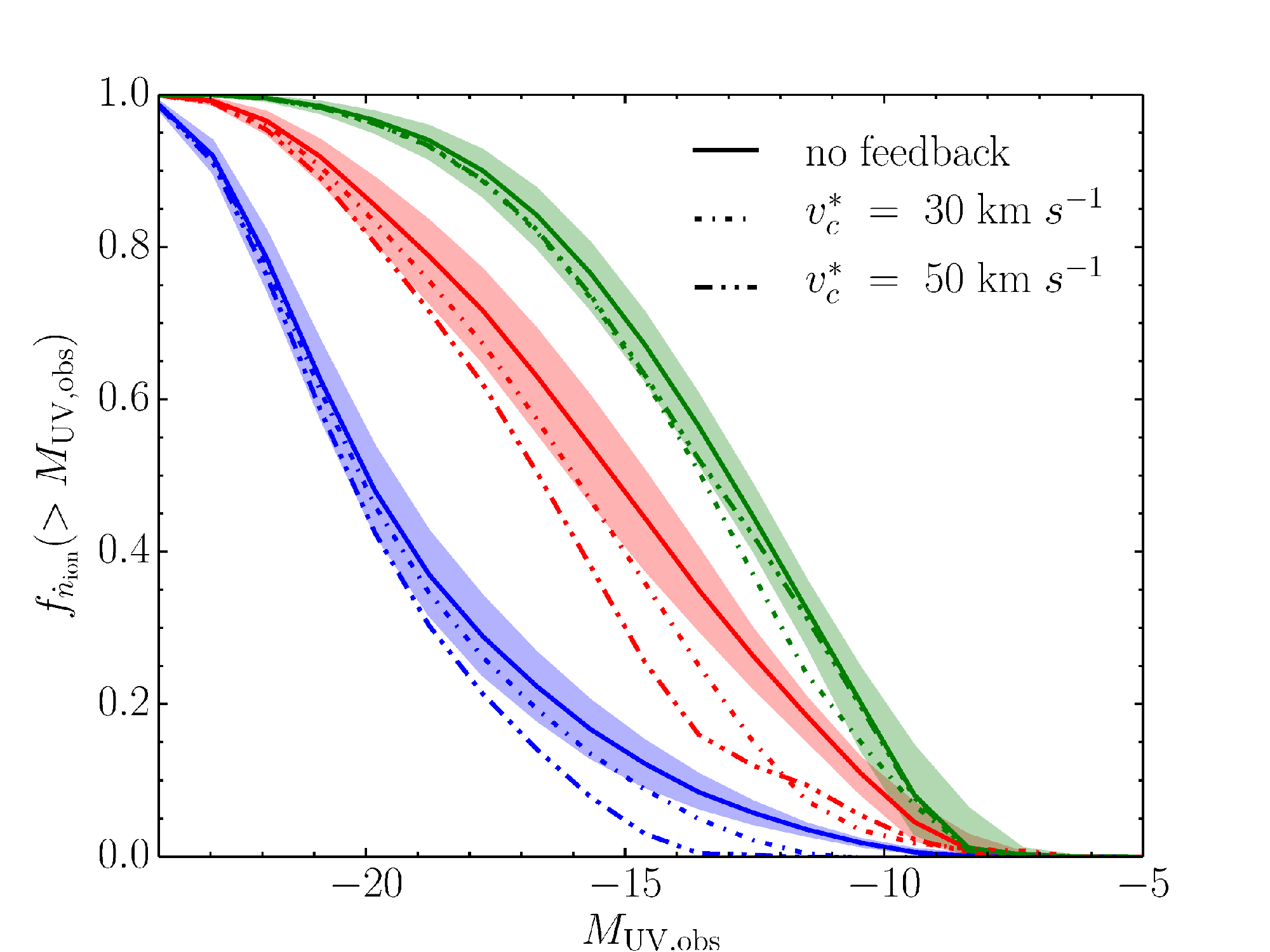}
\caption{Fractional ionizing photon rate from galaxies with absolute UV magnitude above $M_{\rm UV,obs}$, in models with a fixed $f_{\rm esc} = 0.2$, but different $v_c^*$. From top to bottom, each group of curves corresponds to redshift 10, 8, and 5 respectively. To avoid crowding the panel only the 68.3\% confidence intervals of the no-feedback model are shown by shaded regions. 
}
\label{fig_fnion}
}
\end{figure}

Another interesting problem is whether the conclusion that most ionizing photons come from faint galaxies should be revised once  reionization feedback is considered. In Fig. \ref{fig_fnion} we plot the fractional ionizing photon rate produced in galaxies with absolute UV magnitude  above $M_{\rm UV,obs}$, assuming that all galaxies have the same escape fraction of $f_{\rm esc} = 0.2$. It is found that with decreasing redshift, a smaller fraction of ionizing photons comes from fainter galaxies, which is a result of the shallower slope of the LF. When the reionization feedback is taken into account, this fraction is further reduced. However, the qualitative trend remains unchanged. In fact, reionization feedback has significant effects only in the late EoR stages or after EoR, while during the early EoR (e.g. $z\sim10$) faint galaxies always emit the majority of ionizing photons. We note that this conclusion is based on the assumption of a fixed escape fraction for all galaxies. If smaller galaxies have larger escape fraction (e.g. \citealt{2013MNRAS.431.2826F}), they could have been even more dominant compared to the results shown in Fig. \ref{fig_fnion}.

\subsection{Feedback imprints on the LF}

\begin{figure*}
\centering{
\includegraphics[scale=0.4]{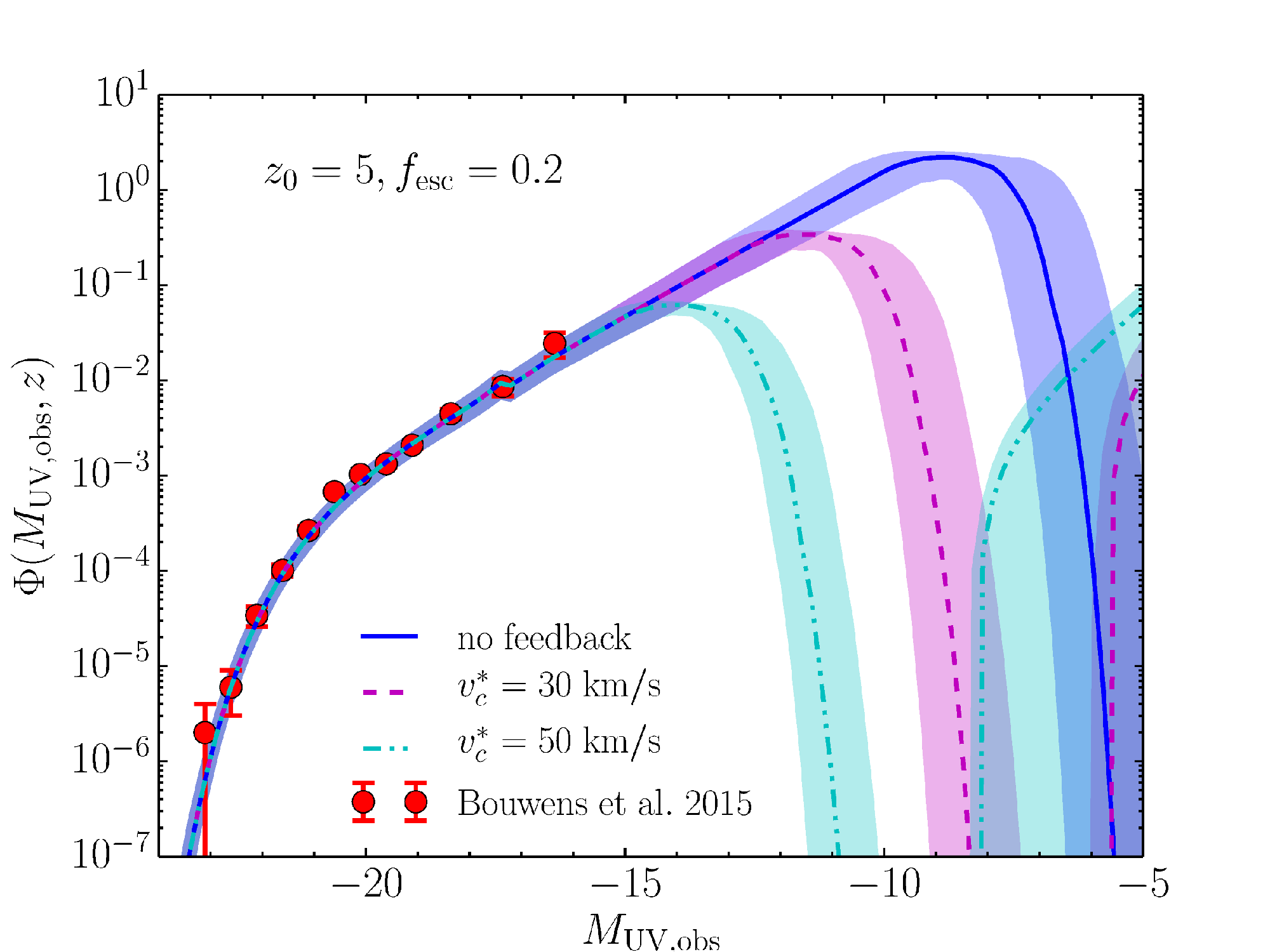}
\includegraphics[scale=0.4]{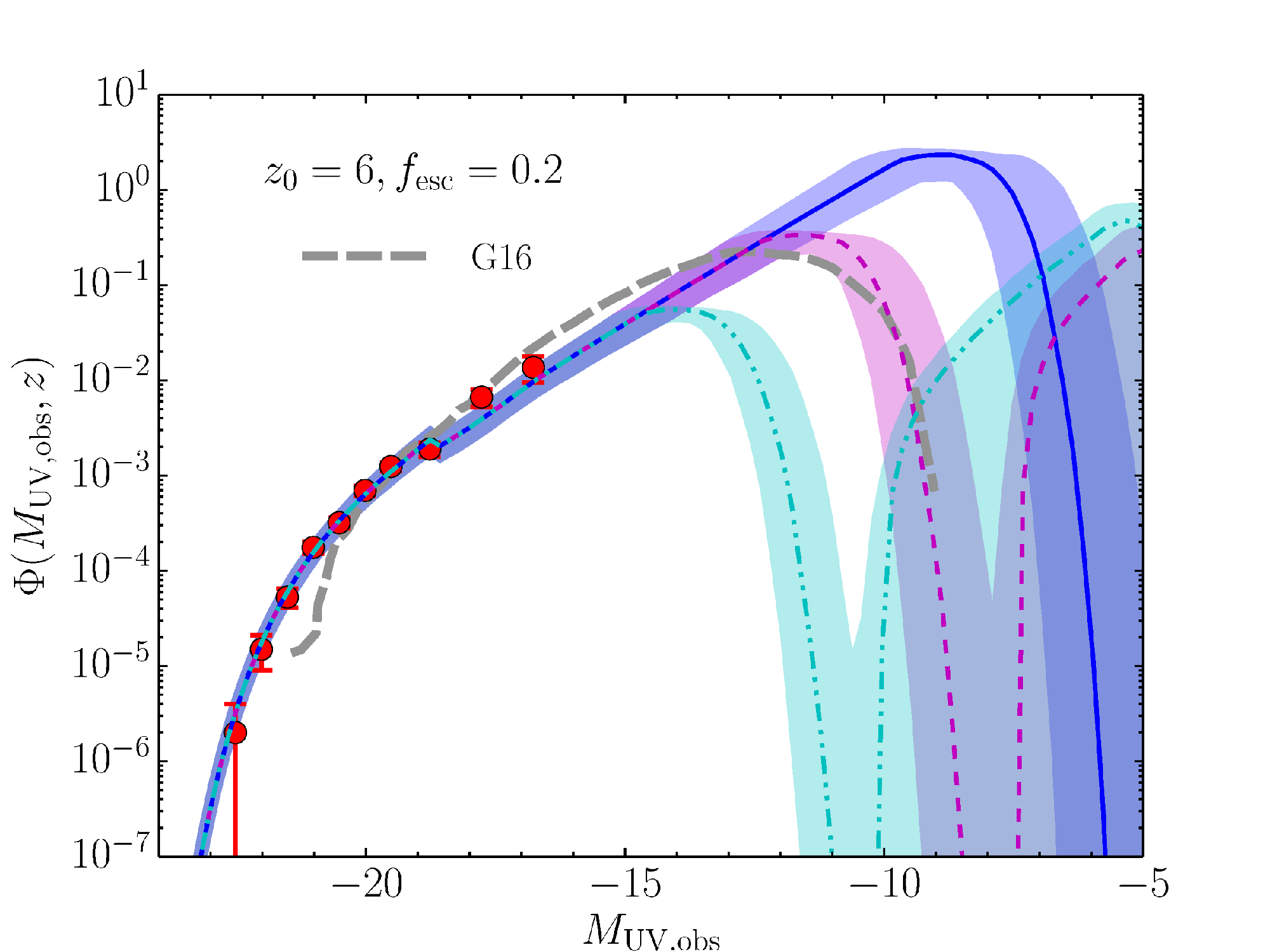}
\includegraphics[scale=0.4]{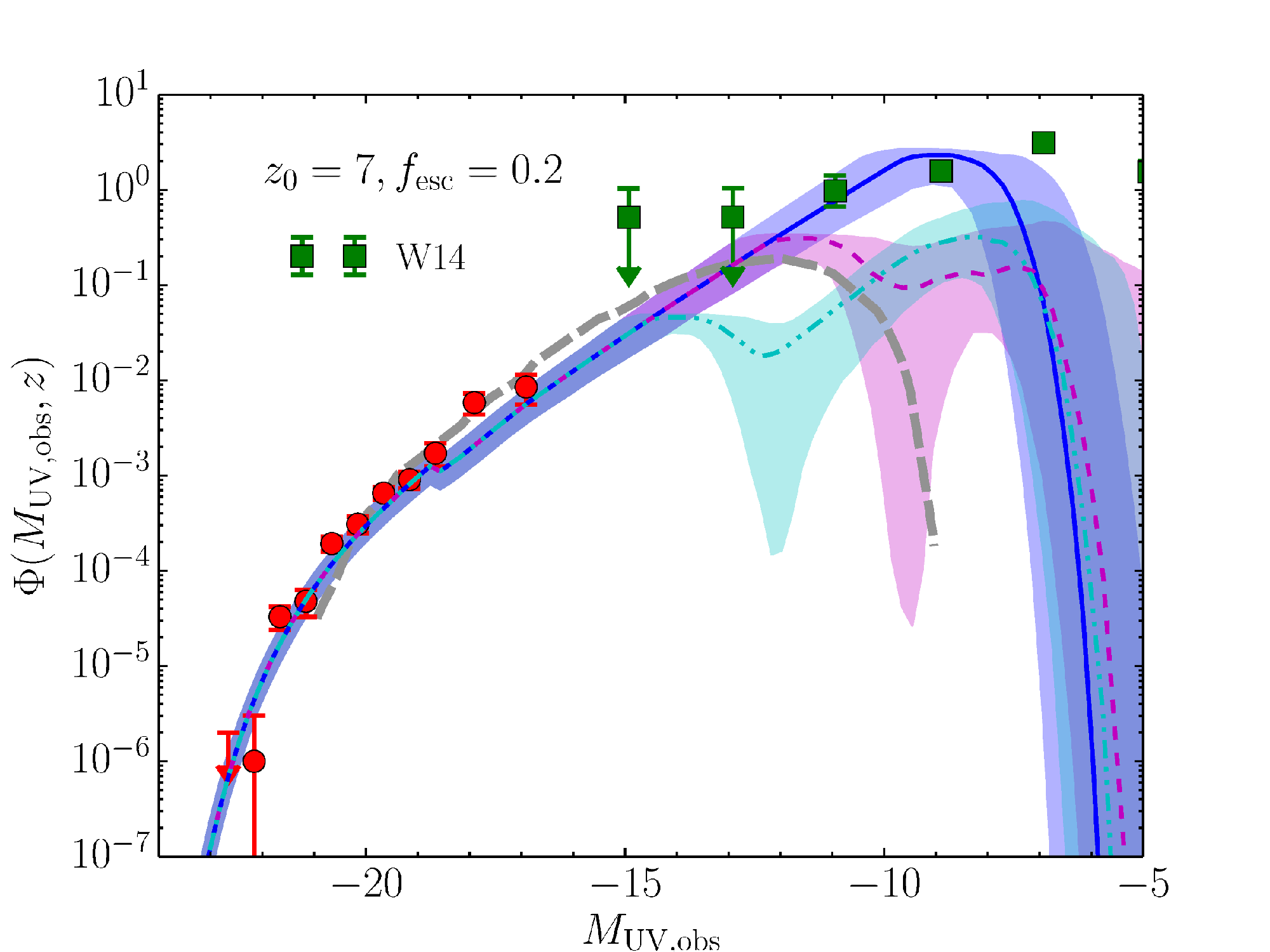}
\includegraphics[scale=0.4]{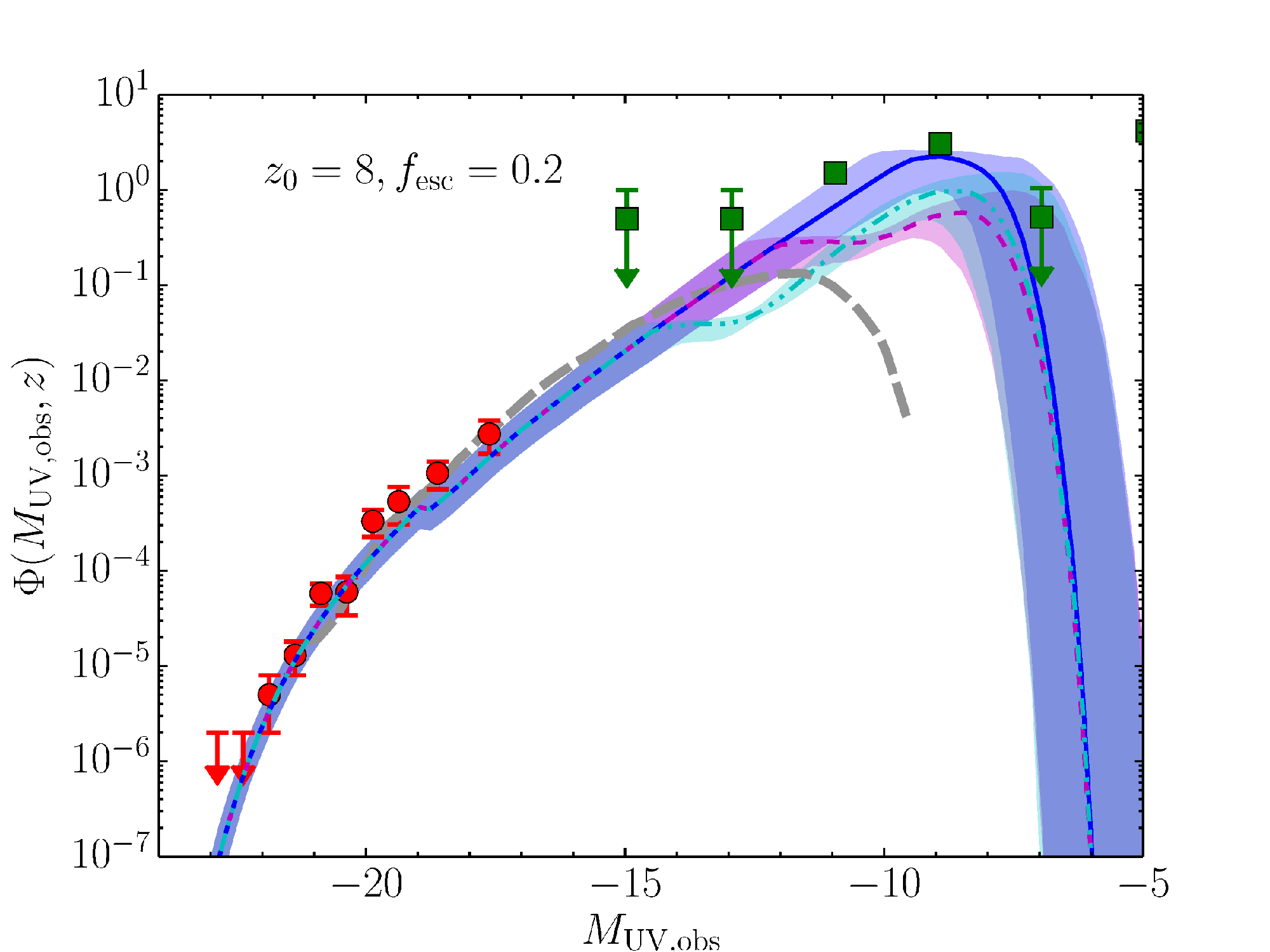}
\caption{The  LFs and uncertainties at redshift 5, 6, 7 and 8 for models with varying $v_c^*$. The shaded regions corresponding each of the models are 
the 68.3\% confidence intervals. For comparison, in panels for redshift 6, 7 and 8 we plot the curves in \citet{2016arXiv160307729G} by thick dashed lines (his uncertainties are not shown); in the panels for redshift 7 and 8, we plot the points in \citet{2014MNRAS.442.2560W} at redshift 7.3 and 8 respectively. Their points with arrows mean there is only one object in the corresponding absolute UV magnitude bin.
}
\label{fig_LF_vc}
}
\end{figure*}

\begin{figure*}
\centering{
\includegraphics[scale=0.4]{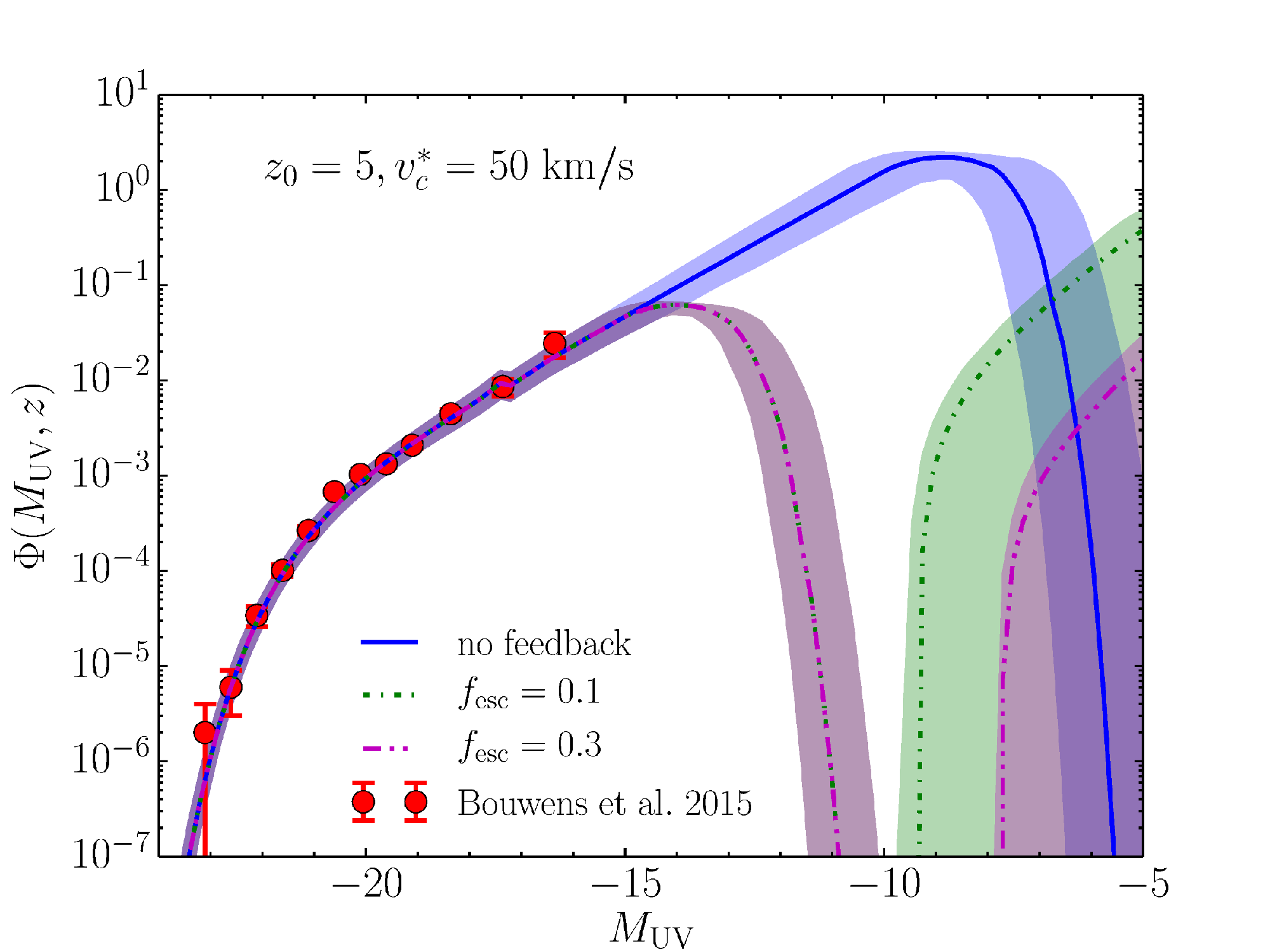}
\includegraphics[scale=0.4]{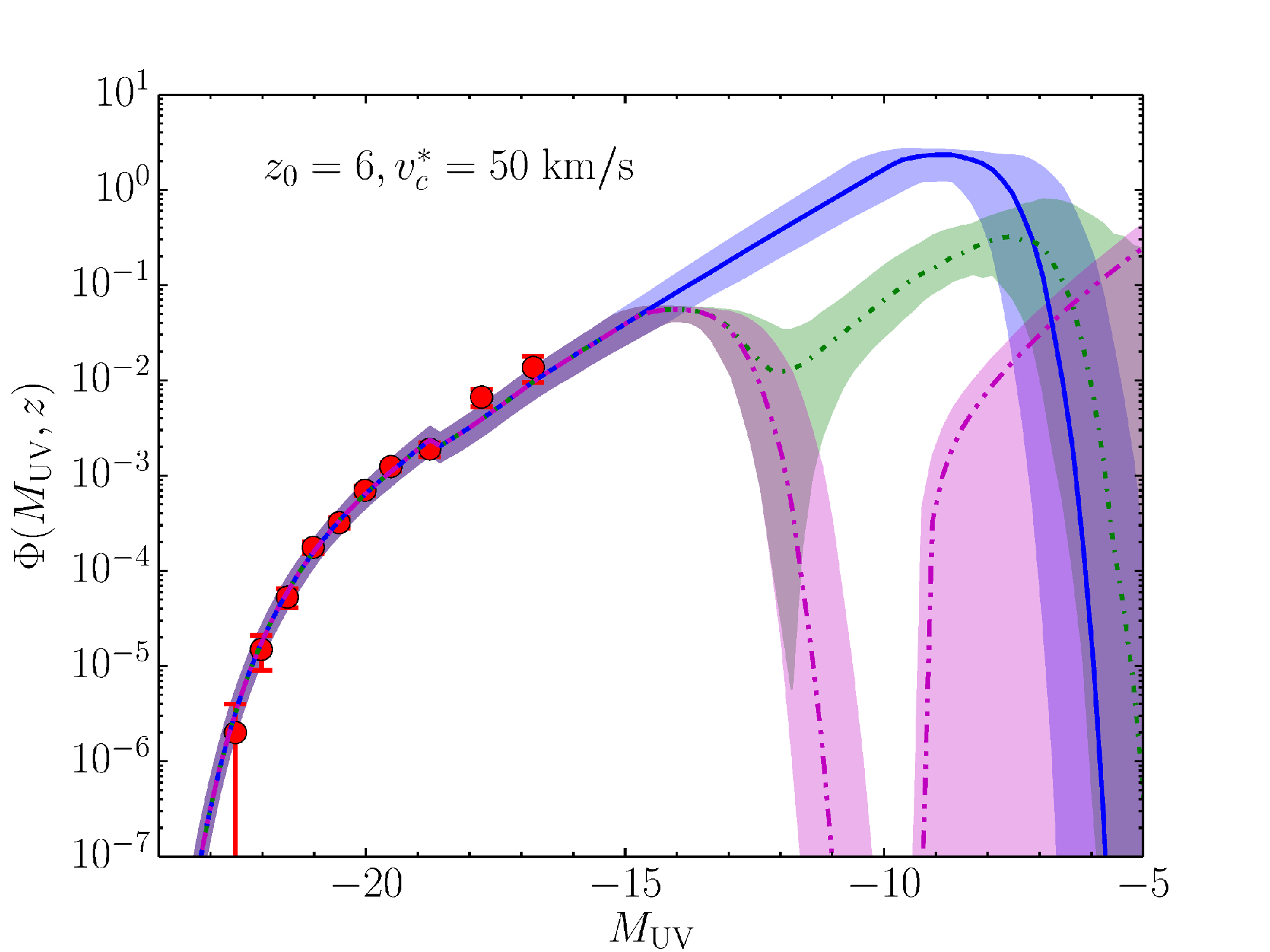}
\includegraphics[scale=0.4]{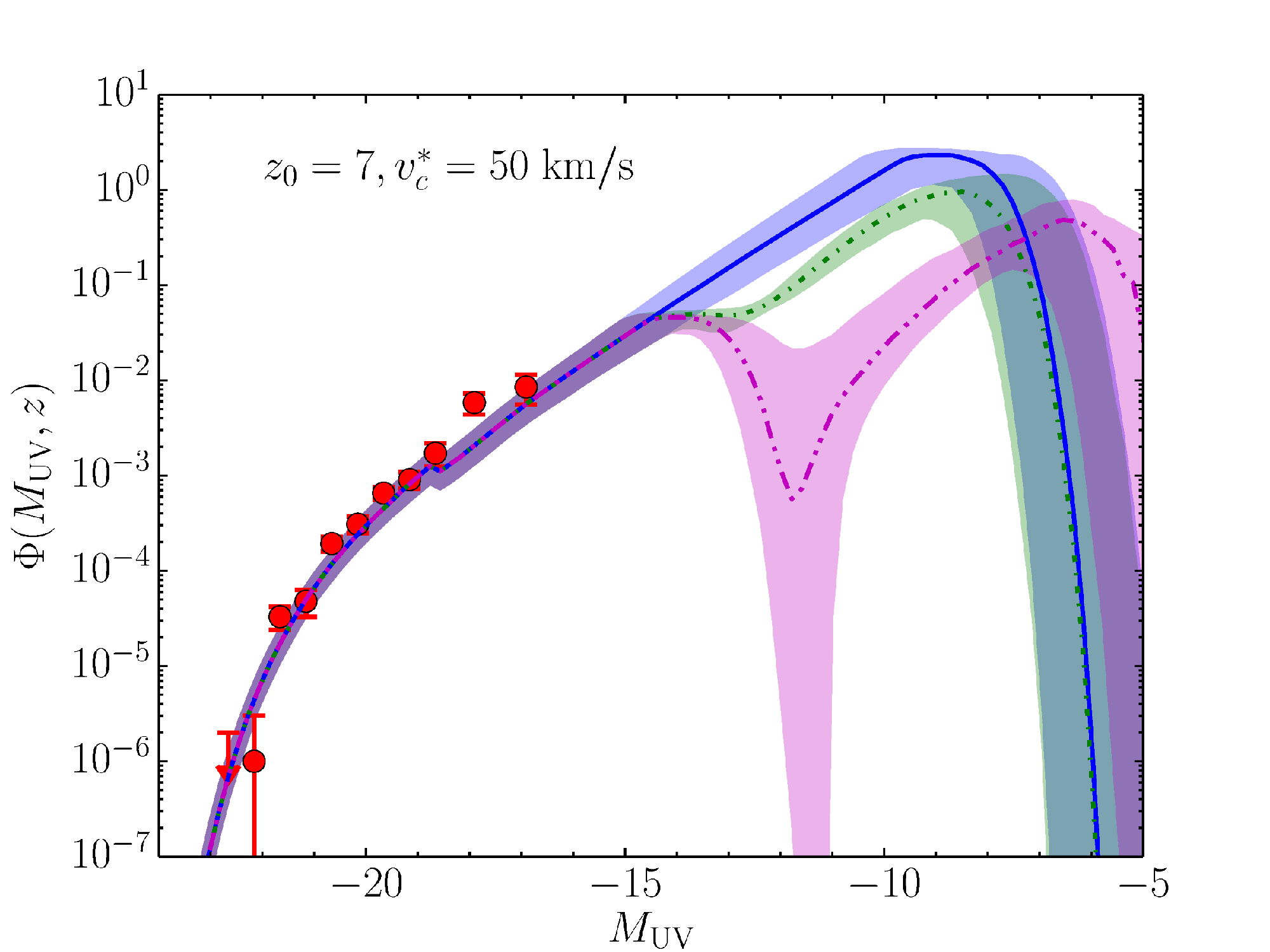}
\includegraphics[scale=0.4]{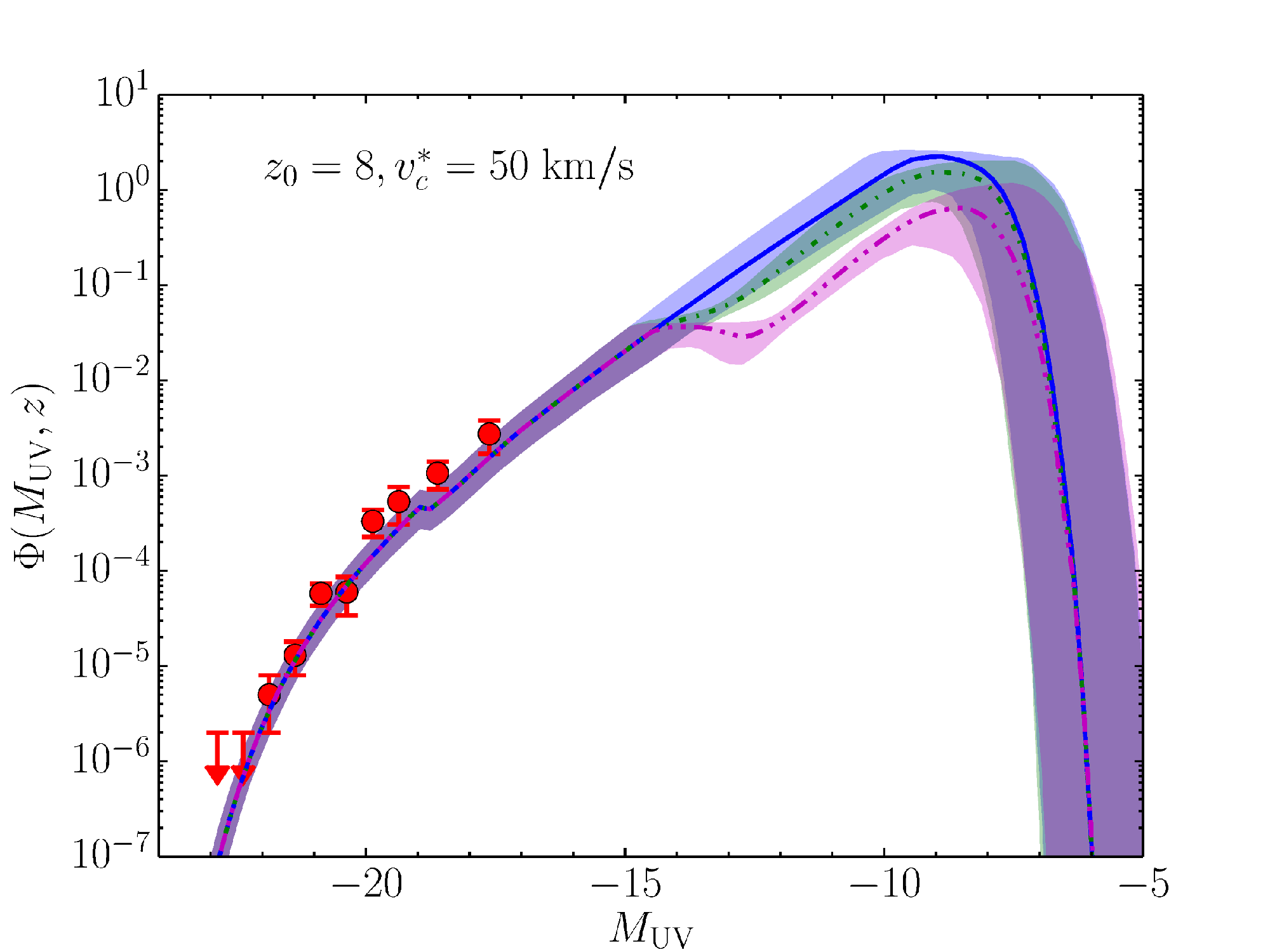}
\caption{Similar to Fig. \ref{fig_LF_vc}, however here $v_c^* = 50 $km s$^{-1}$ while $f_{\rm esc}$ is 0.1 and 0.3 respectively. }
\label{fig_LF_fesc}
}
\end{figure*}

Next we investigate the effect of reionization feedback on the galaxy LF, and how the LF changes with the feedback strength characterized by the threshold $v_c^*$ and $f_{\rm esc}$. We start by fixing again $f_{\rm esc} = 0.2$, and then predict the resulting LFs at $z_0=5, 6 ,7$ and 8 for models with  $v_c^*=30$ and 50 km s$^{-1}$. These LFs and the corresponding uncertainties are shown in Fig. \ref{fig_LF_vc}, where we also plot the LFs for a model with $v_c^*$ corresponding to the atomic-cooling criterion (viral temperature $10^4$ K / $v_c^*=16$ km s$^{-1}$). This no-feedback model is used as a reference. The observational data from \citet{2015ApJ...803...34B} are also plotted. We find that all LFs match observations in the overlapping magnitude ranges, satisfying our first requirement. In addition, we plot in Fig. \ref{fig_LF_fesc} the LFs by adopting $v_c^* = 50$ km s$^{-1}$, while $f_{\rm esc}$ equals 0.1 and 0.3 respectively.

In the no-feedback model, the Schechter parameterization (which can be approximated by $\propto10^{0.4(M_\star - M_{\rm UV,obs}){(\alpha+1)}}$ for $M_{\rm UV,obs} \gg M_\star$) is valid up to a turnover absolute UV magnitude $M_{\rm UV,obs}^* \sim -9$, above which the number of galaxies drops dramatically. This is because the star formation cannot occur in halos below the atomic-cooling criterion, whose typical luminosity corresponds to the turnover UV magnitude $M^*_{\rm UV,obs}\sim-9$. Above this atomic-cooling mass, the decrease in the star formation efficiency in low mass halos is compensated by the increase in the halo number, as seen from Fig. \ref{fig_fsfg}, resulting in a LF with a shallower (albeit still negative) slope with respect to the halo mass function.

The reionization feedback distorts the luminosity - halo mass relation, built from the $f(M_h)$ in Fig. \ref{fig_fsfg} (see detailed discussions in Sec. \ref{sec_galaxies_properties}), for halos with $v_c < v_c^*$, resulting in the complex luminosity distributions of faint galaxies hosted by them. We can generally divide the galaxies into two groups: the ones hosted by halos with $v_c > v_c^*$ and those with $v_c < v_c^*$ respectively. In the former halos, the star formation activity has never been interrupted. For halos with $v_c < v_c^*$, many of them were not able to ignite star formation at all, as they formed in ionized patches; the remaining systems could sustain star formation for some time before being quenched by external ionizing radiation. The latter systems are much fainter than the counterparts that have the same mass assembly history in the reference no-feedback model. As a result the number of galaxies drops dramatically above the new turnover UV magnitude $M_{\rm UV,obs}^*$ roughly corresponding the mean luminosity of halos with $v_c = v_c^*$, namely $M_{\rm UV,obs}^* \approx -12$ ($-15$) for $v_c^* = 30$ (50 km s$^{-1}$).

Galaxies in halos with $v_c < v_c^*$ (with $M_{\rm UV,obs} > M_{\rm UV,obs}^*$) do not totally disappear. Their number drops down above $M_{\rm UV,obs}^*$, but then rises again for higher absolute UV magnitudes. What causes the deficit of galaxies just above $M_{\rm UV,obs}^*$ ? Why, even under strong feedback conditions, the number of faint galaxies remains large (compared to bright galaxies) instead of gradually dropping to zero? The answer to these questions is simple. For halos with $v_c<v_c^*$, only those formed before the end of reionization have the chance to form outside ionized bubbles and start their star formation activity. In the model with $f_{\rm esc} = 0.2$  and $v_c^*=50$ km s$^{-1}$, the reionization completes at $z_{\rm re}=6.6$. Halos formed before $z_{\rm re}$ shine only due to their relic stars, and fade away with time; in halos with $v_c >v_c^*$, most UV radiation is continuously supplied by newly formed stars. As a result, there is a gap in the LFs corresponding to these two (physically distinct) halo populations, and the gap broadens with time.

The number of faint galaxies starts to deviate from the no-feedback model significantly only in the late stages of reionization (EoR lasts for $\sim 300 - 600$ Myr), because earlier on the time stretch is too short to suppress star formation in most halos. By varying the $f_{\rm esc}$, one can change the reionization history. For instance, if we decrease $f_{\rm esc}$ to 0.1 (yielding $z_{\rm re} = 6.0$), as shown in Fig. \ref{fig_LF_fesc}, faint galaxies with $M_{\rm UV,obs} \gsim -9$ could have abundance higher than galaxies as bright as $M_{\rm UV,obs}\sim-18$. 

For comparison, the LFs from hydrodynamical simulations combined with radiative transfer calculation in \citet{2014MNRAS.442.2560W} and in \citet{2016arXiv160307729G} are also shown in the corresponding panels for the same (or very similar) redshifts in Fig. \ref{fig_LF_vc}. We find that at redshift 6 and 7, the \citet{2016arXiv160307729G} LFs are close to our model with $v_c^* = 30~$km s$^{-1}$, however at redshift 8 we predict more faint galaxies above the turnover UV magnitudes than \citet{2016arXiv160307729G}. On the other hand, the \citet{2014MNRAS.442.2560W} LFs are always similar to our no-feedback models. 

\begin{figure*}
\centering{
\includegraphics[scale=0.8]{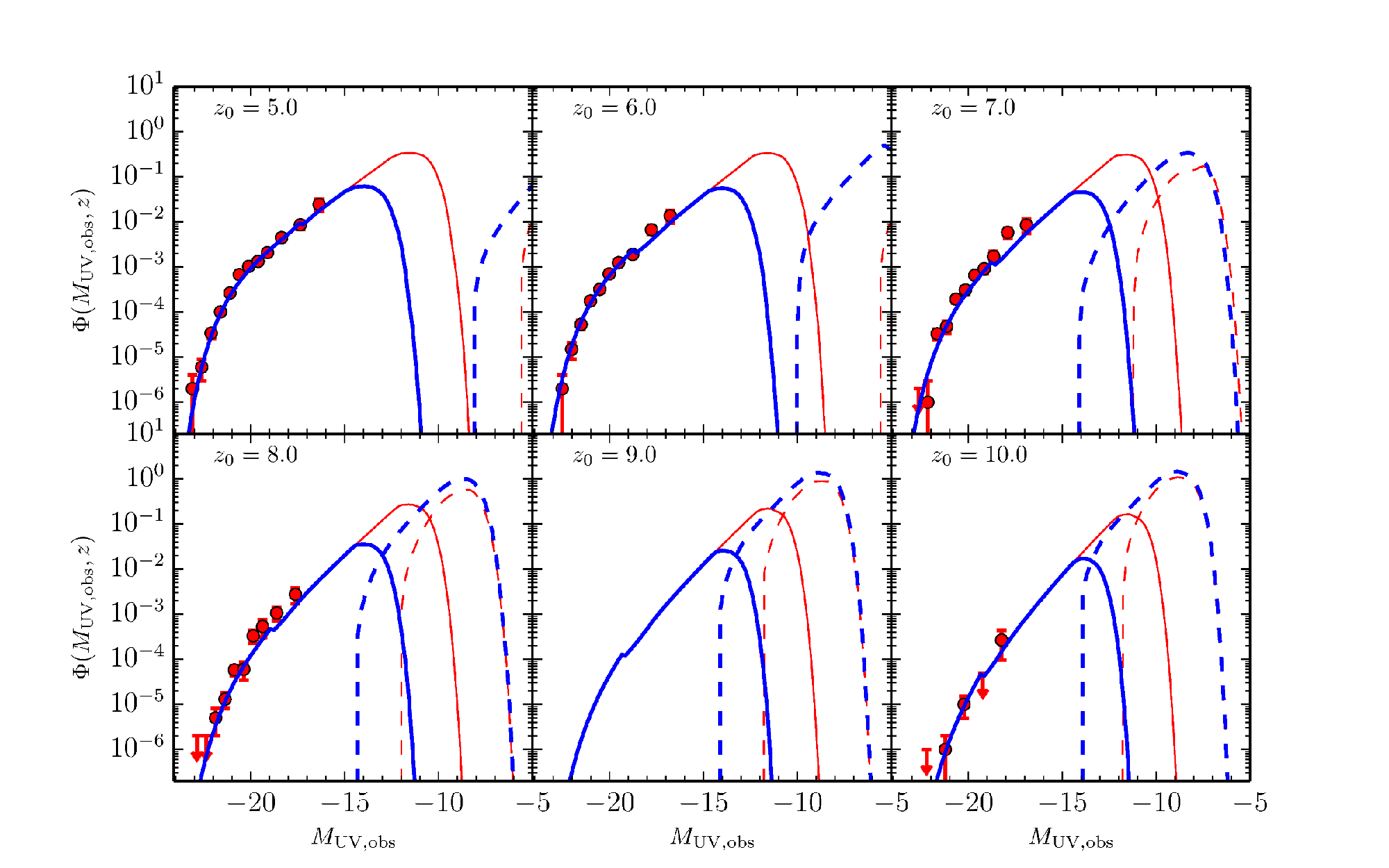}
\caption{The contributions from halos with $v_c < v_c^*$ (dashed) and $v_c > v_c^*$ (solid) to the predicted LFs at different redshifts for models with $v_c^* = 30$ (thin) and 50 km s$^{-1}$ (thick) respectively. To highlight the difference between these two components, uncertainties are not shown.}
\label{fig_LF_two_components}
}
\end{figure*}

To get a deeper insight, we  plot separately the LFs of halos with $v_c >v_c^*$ and with $v_c <v_c^*$ at different redshifts in Fig. \ref{fig_LF_two_components}, for $v_c^* = 30$ and 50 km s$^{-1}$, respectively. Also shown are the available observational data from \citet{2015ApJ...803...34B}. From the figure we clearly see how reionization feedback gradually separates the two components inducing an increasing deviation from the no-feedback LF reference model.  It is worthwhile noting that although we just simply adopt a constant circular velocity threshold as a star formation quenching criterion, the LFs have rather complex behaviors at the faintest end. From $z\sim8$ to $z\sim5$, the abundance of galaxies with $M_{\rm UV,obs} > M_{\rm UV,obs}^*$ evolves fast, and these faint galaxies hosted by halos with $v_c < v_c^*$  become EoR relics (see e.g. \citealt{2005ApJ...629..259R,Salvadori09, 2015MNRAS.450.4207B}). 

The above discussions are only concerned with the $f(M_h)$ calibrated by using best-fitted Schechter parameters. Considering the uncertainties, predictions on the number of galaxies at the faint-end ($M_{\rm UV,obs}\gg M_{\rm UV,obs}^*$) are rather uncertain, as shown by shaded regions in each panel of Fig. \ref{fig_LF_vc} and Fig. \ref{fig_LF_fesc}. On the other hand, the observed abundance of ultra-faint galaxies can put tight limits on the star formation efficiency of small halos.

\subsection{Feedback imprints on galaxy properties}\label{sec_galaxies_properties}

\begin{figure}
\includegraphics[scale=0.45]{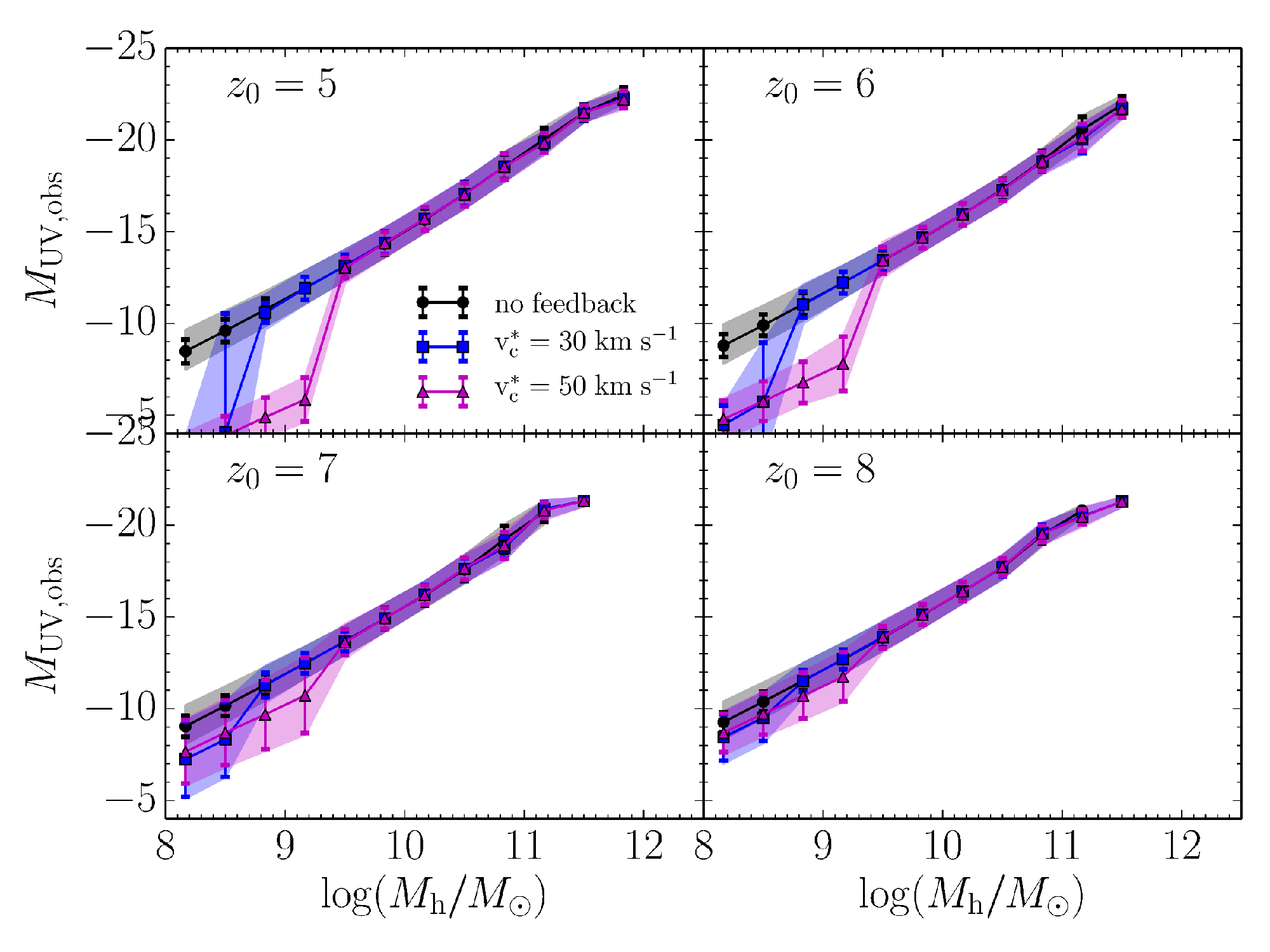}
 \caption{The $M_{\rm UV,obs} - M_{\rm h}$ relations for halos at redshifts 5, 6, 7 and 8 in models with $v_c^* = 30$ and 50 km s$^{-1}$, and in no-feedback model, respectively. Errorbars are the intrinsic scatters of the absolute UV magnitudes in each halo mass bin, while the shaded regions are the full uncertainties including both the intrinsic scatters and the uncertainties in calibrating the $f(M_{\rm h})$.
}
\label{fig_MUV-Mh}
\end{figure}

We further investigate the imprints of reionization feedback on galaxies properties. The first is the $M_{\rm UV,obs}-M_{\rm h}$ relation, which critically concerns the galaxy LFs discussed above. As different halo mass assembly histories introduce an intrinsic scatter in this relation, in order to derive the scatter, it is more appropriate to generate Monte Carlo random samples. The samples are generated by using the probability distribution of the (a) halo mass (given by  the halo mass function), (b) formation time (from Eq. \ref{pw} for a given halo mass $M_{\rm h}$, and redshift $z_0$), and (c) quenching time (using $\propto d{\mathcal P}_b/dz_q$ and $z_0 < z_q <z_f$) obtained above. In Fig. \ref{fig_MUV-Mh} we show the results at redshifts 5, 6 ,7 and 8, for the no-feedback reference model, and for $v_c^* = 30$ and 50  km s$^{-1}$. All models have $f_{\rm esc} = 0.2$. 
We use the error bars to represent the intrinsic scatter of the absolute UV magnitudes due to different mass assembly histories of halos in the same mass bin and due to different reionization imprints on them, and shaded regions to represent the full uncertainties considering both the intrinsic scatters and the uncertainties in calibrating the star formation efficiency. The reionization feedback decreases the mean luminosity of halos with $v_c < v_c^*$ at the same time increasing the scatter, imprinting an ankle-knee feature in the relation.

\begin{figure}
\centering{
\includegraphics[scale=0.4]{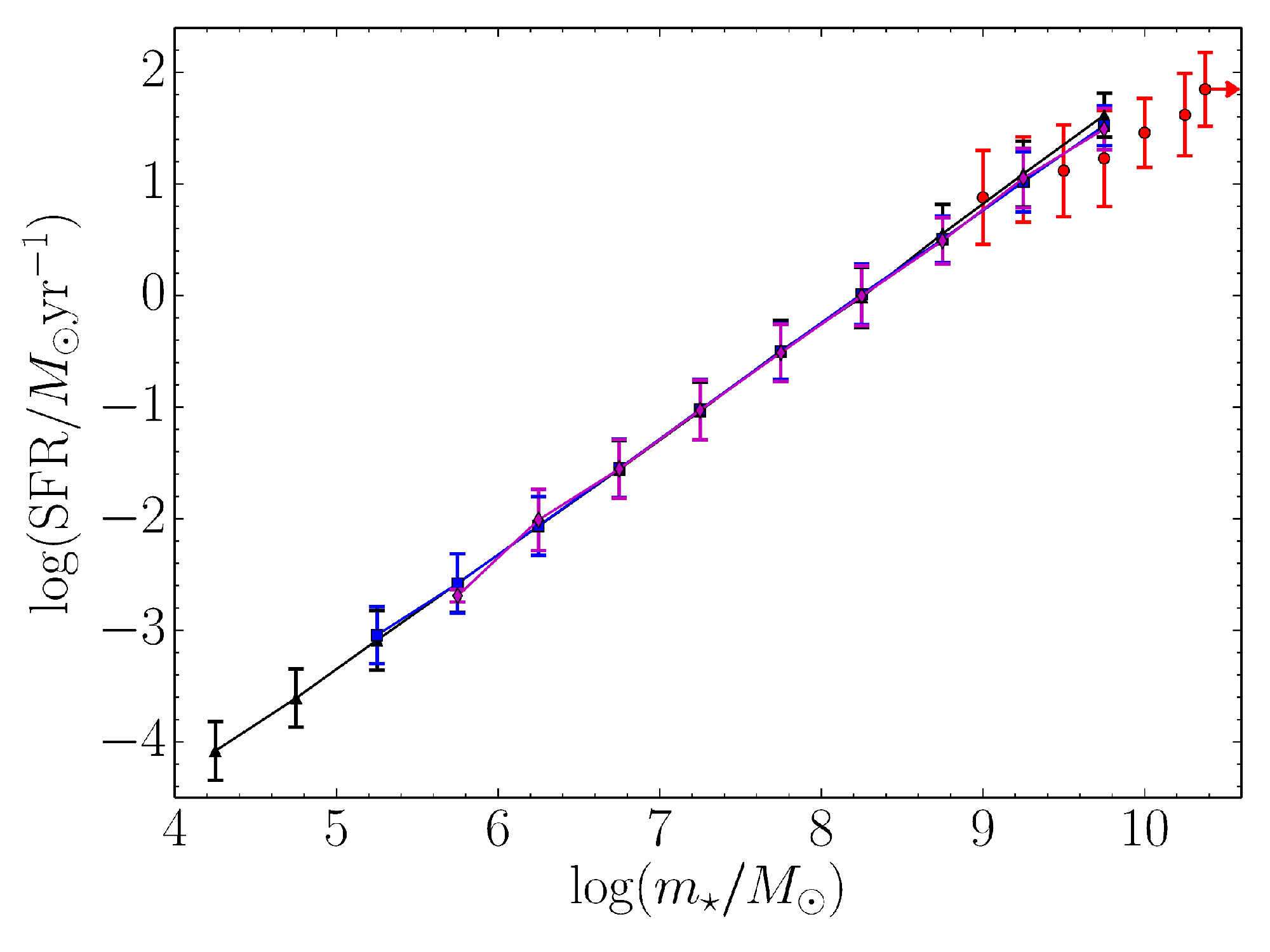}
\caption{The SFR vs. $m_\star$ and the corresponding uncertainties at redshift 5 for three models respectively. For comparison we also plot the observed relation in \citet{2015ApJ...799..183S}. We only show the intrinsic scatters because in the SFR - $m_\star$ relation the uncertainties due to calibrating star formation efficiency is very small and scarcely visible.
}
\label{fig_SFR_Ms}
}
\end{figure}

We also extract the SFR - $m_*$ and $t_*$ - $M_{\rm UV,obs}$ relations from the above Monte Carlo samples. The SFR - $m_\star$ relation at $z_0=5$ is shown in Fig. \ref{fig_SFR_Ms}, and compared with data from \citet{2015ApJ...799..183S}. Our predicted SFR - $m_\star$ relations agree well with observations in the overlapping $m_\star$ range. Discrepancies in the SFR - $m_\star$ relations of the three models are modest at $m_\star \gsim 10^6~M_\odot$; however, in the $v_c^* = 30$ (50) km s$^{-1}$ models star formation in all galaxies with $m_\star \lsim 10^5 (10^6)~M_\odot$ has already been quenched before redshift 5. The SFR - $m_\star$ relations at higher redshifts have a similar trend, except that the amplitudes increase by about 0.1 dex per redshift.

\begin{figure}
\includegraphics[scale=0.45]{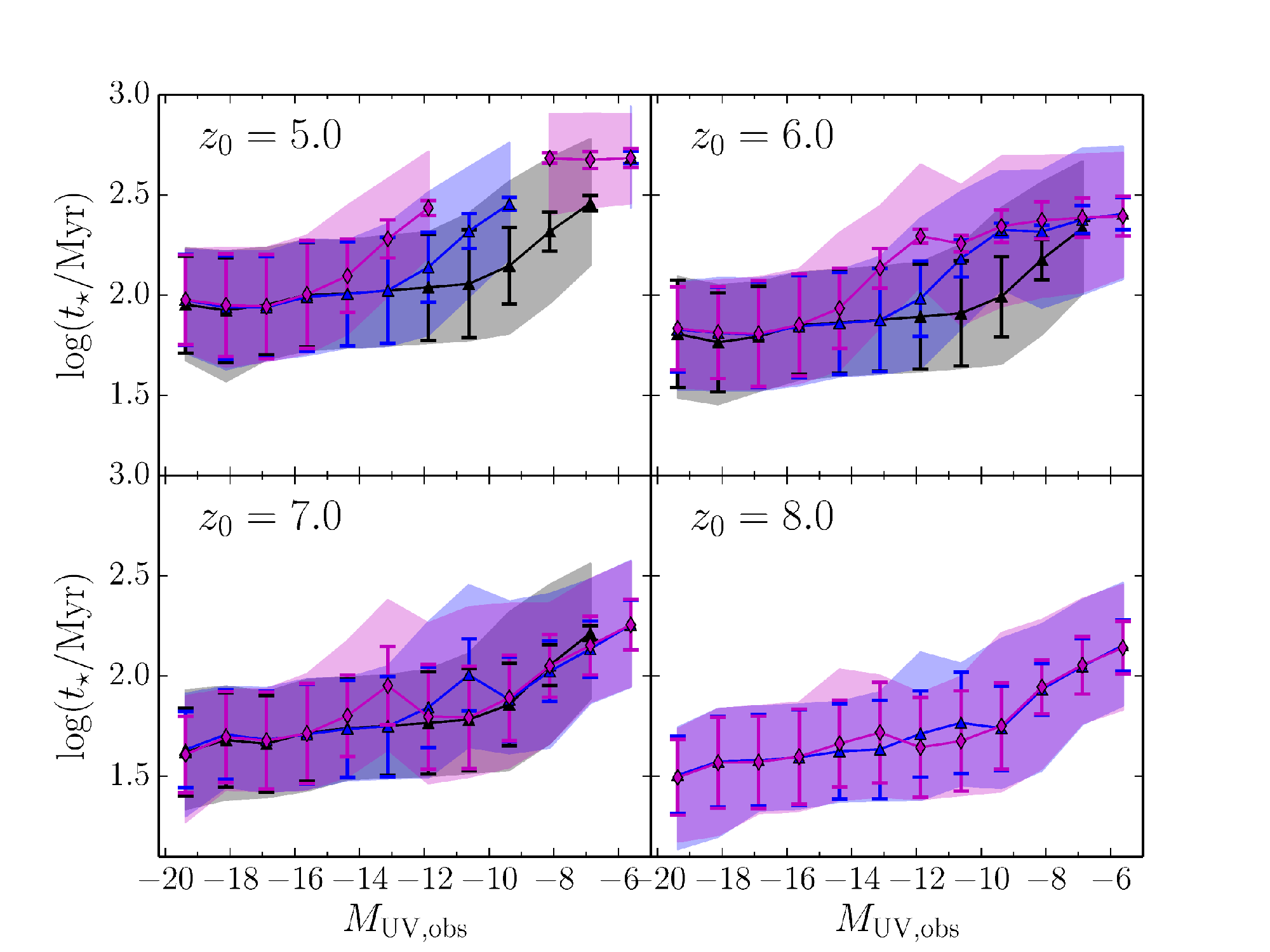}
\caption{ The $t_\star$ - $M_{\rm UV,obs}$ relations and the corresponding uncertainties at redshifts 5, 6 ,7 and 8 respectively. Same to Fig. \ref{fig_MUV-Mh}, errorbars mean the intrinsic scatters while the shaded regions mean the full uncertainties.}
\label{fig_ts_MUV}
\end{figure}

The analogous stellar age ($t_\star$) vs. $M_{\rm UV,obs}$ relation is plotted in Fig. \ref{fig_ts_MUV}. We find that at redshift 5 galaxies with $M_{\rm UV,obs} \lsim -16$ have mean stellar age $\sim100$~Myr. Above this absolute UV magnitude models start to diverge, with stronger feedback models predicting relatively older galaxies. By looking at the $t_\star$ - $M_{\rm UV,obs}$ relations  it is easier to distinguish models in the range $-13 \lsim M_{\rm UV,obs}\lsim -8$. Therefore, stellar age measurements of galaxies in this range could be used as a probe of reionization feedback strength. At higher redshifts stars are typically younger: for example at $z=8$ galaxies with $M_{\rm UV,obs} < -10$ are about 30 Myr old.

\subsection{Feedback imprints on galaxy counts}
\begin{figure}
\centering{
\includegraphics[scale=0.4]{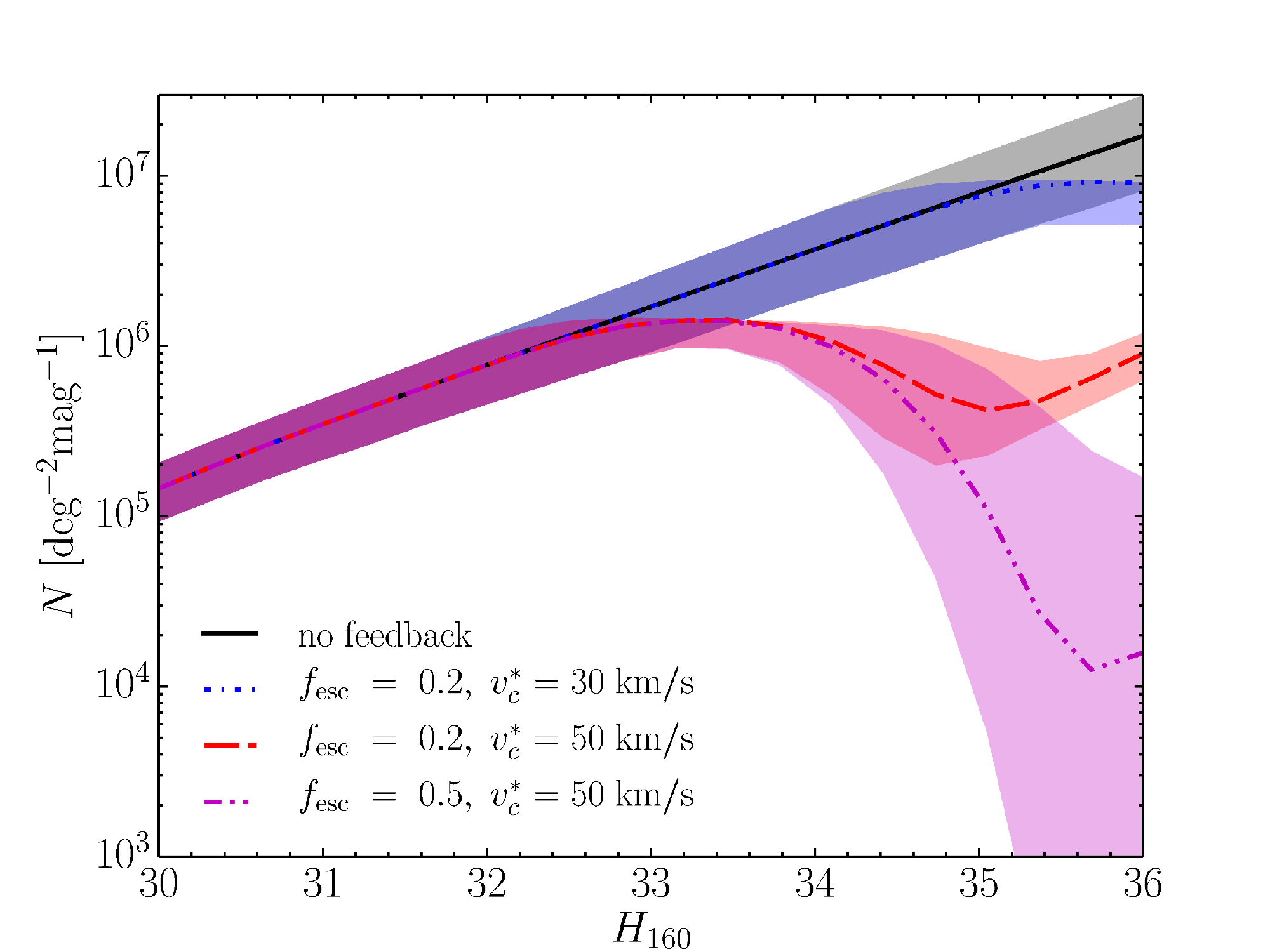}
\caption{The predicted number counts of galaxies with $z=5 - 8$ in models with different parameters.}
\label{fig_Nm}
}
\end{figure}

We have pointed that our model predicts a Thomson optical depth $\tau$ consistent with the {\tt Planck} constraints in a wide $v_c^*$ range. Therefore this indirect observation is not very helpful in discriminating models with different $v_c^*$.  However, at the faintest magnitudes different models predict substantially different galaxy number counts in a given redshift range from $z_1$ to $z_2$, 
 \begin{equation}
N(H_{160}) =\int_{z_1}^{z_2} r^2\Phi(M_{\rm UV,obs},z)\frac{dr}{dz}dz,
\end{equation}
where $H_{160}$ is the apparent magnitude observed at 1.6~$\mu$m, corresponding to rest-frame luminosity at $1.6/(1+z)$~$\mu$m. This rest-frame luminosity is converted to the absolute UV magnitude at 1600~\AA~using the $l_\nu(\Delta t)$ as in Eq. (\ref{Lz}), but at fixed $\Delta t = 100$~Myr for convenience. Hence, number count observations could directly put constraints on $v_c^*$, as shown in Fig. \ref{fig_Nm} for galaxies in the redshift range 5 - 8. Among these ultra-faint galaxies a substantial fraction of them are located in $z=7-8$, due to the steeper slope of the LF at $z\sim8$ compared to at $z\sim5$. For example, we check that for the number count of no-feedback model shown in Fig. \ref{fig_Nm}, at $H_{160}\sim30$ about $\sim16$\% is in $z=7-8$, at $H_{160}\sim36$ this fraction is $\sim25$\%. Existing and/or forthcoming galaxy surveys are unlikely to reach the very deep limiting magnitudes required. However, if gravitational lensing (e.g. \citealt{2014MNRAS.443L..20Y}) can be exploited, it is possible to detect a handful of ultra-faint galaxies, that would allow to put tight constraints on feedback strength.  The results of investigations using two Frontier Fields clusters are presented in \citet{2016ApJ...823L..40C}.

\section{Conclusions}\label{conclusions} 
The star formation activity in small halos with shallow gravitational potential well is easily quenched by external ionizing flux from nearby sources and/or an ionizing radiation background. Such quenching effect might play a significant role in shaping the reionization history, when more and more galaxies formed in/entered into ionized bubbles whose size keeps growing throughout the EoR. Thereby the LF of such faint galaxies provides key information on the interplay between the reionization process and its driving sources.

We have investigated the LF of faint galaxies during the EoR by including the above reionization feedback in an analytical model. 
The model derives the LF from halo mass function by constructing luminosity - halo mass relations from observationally-calibrated star formation efficiency and halo mass assembly history. 
Reionization feedback effects are included by adopting a constant threshold circular velocity, $v_c^*$, below which the star formation of halos located in ionized bubbles is quenched. 
We computed the LFs for models with different $f_{\rm esc}$ and $v_c^*$ values, and found that:
\begin{itemize}
\item If reionization feedback is neglected, the power-law Schechter parameterization characterizing the faint-end of the LF remains valid up to $M^*_{\rm UV,obs}\sim -9$ (corresponding to the atomic-cooling halos mass, see Appendix \ref{SN} for an estimate of supernova explosion effects). Above this absolute UV magnitude the number density of galaxies drops dramatically.  

\item  When feedback is included, small halos ($v_c < v_c^*$) in ionized bubbles fail to collect enough gas to ignite/sustain their star formation. The reionization history, constrained by the {\tt Planck} electron scattering optical depth, is insensitive to $v_c^*$.  

\item For strong feedback, i.e. $v_c^* = 50$~km s$^{-1}$, the LF deviates from the Schechter function above $M^*_{\rm UV,obs} \sim -15$, slightly below the detection limit of current surveys of blank fields at $z\sim5$ \citep{2015ApJ...803...34B}. Hence, we expect that upcoming observations will obtain important constraints on $v_c^*$.

\item In addition, even for strong feedback, the LF may rise again at luminosities fainter than $M^*_{\rm UV,obs} $ as a result of the interplay between reionization process and galaxy formation. 

\item We also pointed out that the $t_\star$ - $M_{\rm UV}$ relation might be used as a powerful probe of reionization feedback strength. In models with stronger reionization feedback, stars in galaxies with $-13 \lsim M_{\rm UV,obs} \lsim -8$ are typically older. Other constraints on $f_{\rm esc}$ and $v_c^*$ in our model, can come from galaxy number count data, particularly from those exploiting gravitational lensing magnification. 
\end{itemize}
 
Our model contains some necessary simplifications and assumptions. The most relevant one is perhaps the use of a constant circular velocity threshold as the criterion for quenching star formation, and its treatment 
as a free parameter independent of $f_{\rm esc}$. This is a standard assumption in the literature and is very convenient when performing analytical calculations. Instead, detailed simulations (e.g. \citealt{2013MNRAS.432L..51S}) pointed out that in the presence of reionization feedback, the gas fraction decreases gradually as the halo mass decreases, following a relation $2^{-M_c/M_{\rm h}}$, where $M_c$ is a critical halo mass. We believe that this effect would make the LF smoother around the turnover point and may 
result in more faint galaxies. However, we do not expect significant changes in the basic trend of the galaxy LF found here.

In our work, we take the model parameters that give predictions consistent with the direct observations of the LFs at $z > 5$ in the blank fields. There are alternative ways to investigate the reionization feedback and constrain the high-$z$ LFs using indirect observations. At intermediate redshifts ($2 \lsim  z \lsim 5$) the IGM is fully ionized and a global ionizing UV background is in place. Hence the radiative feedback effects should be maximal. The known existence of faint galaxies in these epochs implies the existence of even more fainter galaxies in the EoR.

For example, in \citet{2014ApJ...780..143A} the observations of galaxies at $z\sim2$ confirm the validity of the Schechter formula down to absolute UV magnitude $\sim-13$. These results show that such faint galaxies  support active star formation well after the EoR. As a consequence, it is likely that $v_c^* \lsim 50$ km s$^{-1}$ if we incorporate this information into our model.

In addition, the number of ultra-faint satellites in the Local Group also put constraints on the faint-end of the EoR LF once used in combination with their merger tree history, see \citet{2014ApJ...794L...3W,2014MNRAS.443L..44B,2015MNRAS.453.1503B}. Finally, the rate of high-$z$ GRBs is another probe of the abundance of ultra-faint galaxies in the EoR \citep{2012ApJ...749L..38T}.  Here, we checked that distinguishing models with different $v_c^*$ requires very high precision measurements of the star formation rate density (SFRD). Even at $z\sim5$ the SFRD difference between the no-feedback model and the $v_c^* = 50$ km s$^{-1}$ is only about 10\%, i.e. much smaller than the current precision of the SFRD derived from GRB observations. Nevertheless, all these alternative techniques nicely complement investigations, as the one presented here, based on direct LFs or number count data.

\section*{Acknowledgements}
YX is supported by the NSFC grant 11303034, and the Young Researcher Grant of 
National Astronomical Observatories, Chinese Academy of Sciences.


\appendix

\section{Supernova feedback} \label{SN}
The star formation efficiency is calibrated by assuming that the Schechter parameterization for LFs always holds down to the atomic-cooling halo mass. This might not be true for small halos in which gas could be  totally blown away by supernova explosion, therefore in this section we have a check on it. Halos can sustain the continuous star formation mode if the energy deposited by supernova explosions does not exceed the gravitational binding energy of the halo,
\begin{equation}
m_\star \eta_{\rm SN}E_{\rm SN} < \frac{1}{2}M_{\rm h}f_{\rm g} v^2_{\rm esc}=M_{\rm h}f_{\rm g}v^2_c,
\end{equation} 
where $E_{\rm SN} $ is the energy released by supernova per stellar mass and $\eta_{\rm SN} \sim 0.1$ \citep{2014MNRAS.440.2498P} is the fraction that this energy goes into the gas, $f_{\rm g}$ is the gas fraction and $v_{\rm esc}=\sqrt{2}v_c$ is the escape velocity. In our case $m_\star \approx M_{\rm h }f(M_{\rm h})/2$, we have  
\begin{equation}
f(M_{\rm h})\lsim \frac{2f_{\rm g}v^2_c}{\eta_{\rm SN}E_{\rm SN}}.
\label{fupper}
\end{equation}
If the upper limit at the right hand side is smaller than the calibrated star formation efficiency the halo has to adjust itself to have the new star formation efficiency satisfies the Eq. ({\ref{fupper}}) \citep{2014MNRAS.445.2545D}. Simply assuming $f_{\rm g}=\Omega_b/\Omega_m$, and taking $E_{\rm SN}$ from the outputs of {\tt Starburst99}\footnote{http://www.stsci.edu/science/starburst99/docs/default.htm} (\citealt{1999ApJS..123....3L,2005ApJ...621..695V,2010ApJS..189..309L}, for consistence reason we also replace the $l_\nu(\Delta t)$ in Eq. \ref{Lz} with the one from {\tt Starburst99} in this check),
we check that for the Salpeter IMF with the mass ranges of 0.1 - 100 $M_\odot$, $1 - 100 M_\odot$  and $3 - 150 M_\odot$,  the Eq. (\ref{fupper}) always holds for halo mass above the atomic-cooling mass, as long as $\eta_{\rm SN}\lsim 0.4$ when the metallicity is $0.02~Z_\odot$, and as long as $\eta_{\rm SN}\lsim0.3$ when the metallicity is $2~Z_\odot$. We therefore believe that our assumption is safe enough. This is because in our work for any given metallicity and IMF, we calibrate the corresponding $f(M_{\rm h})$ to reproduce the observed LF at $z_0 \sim 5$. In the IMF models with more massive stars, both the $E_{\rm SN}$ and the $l_\nu(\Delta t)$ are higher. The higher $l_\nu(\Delta t)$ consequently results in a smaller $f(M_{\rm h})$, limiting the total energy released by supernovae $\propto f(M_{\rm h}) \eta_{\rm SN} E_{\rm SN}$, so the host halos could still hold the star formation activity. The effects by Pop III stars are not considered here, as pointed by numerical simulations \citep{2007MNRAS.382..945T,2014MNRAS.440.2498P} Pop III stars are negligible in halos above the atomic-cooling criterion.
 
\section{  ${\mathcal P}_{\rm b}$ vs. $Q_{\rm HII}$} \label{PBvsQ}

\begin{figure}
\centering{
\includegraphics[scale=0.4]{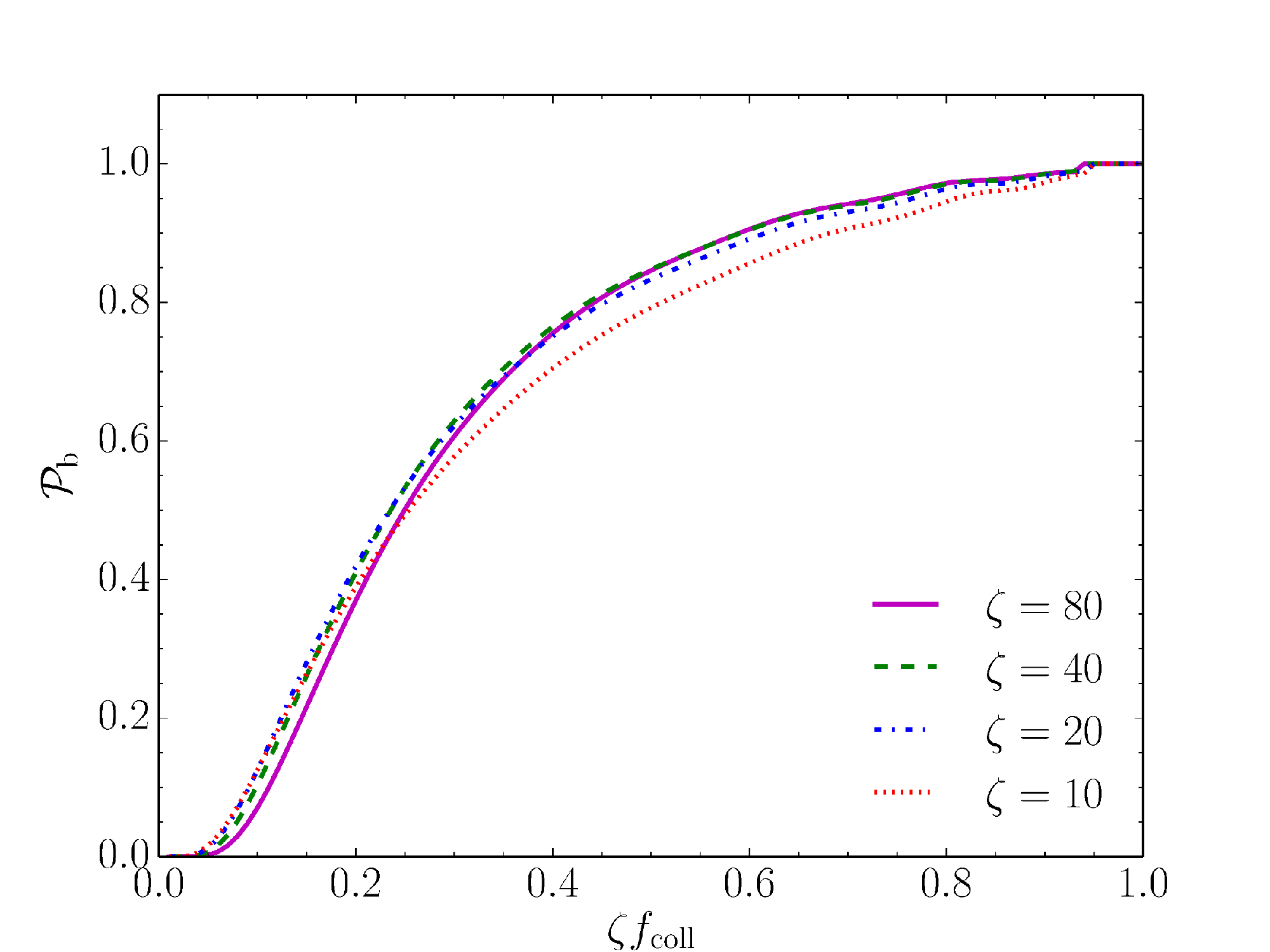}
\caption{The ${\mathcal P}_{\rm b}$ for $M_{\rm h} = 10^9~M_\odot$  as a function of $\zeta f_{\rm coll}$ for different $\zeta$. }
\label{fig_Pb}
}
\end{figure}
We integrate Eq. (4) in \citet{2004MNRAS.354..695F} to calculate the probability that a halo sits in ionized bubbles above a minimum size. This is basically an application of the ``bubble model" scenario discussed in \cite{2004ApJ...613....1F}. The minimum bubble size is set by requiring that a target halo has at least one neighbor with $v_c>v_c^*$ within the bubble radius. Namely, we have the two-point halo correlation function:
\begin{equation}
\xi_{h}(m_1,m_2,r,z)=\xi(r,z)b(m_1,z)b(m_2,z),
\end{equation}
where $\xi$ is the matter two-point correlation function and $b$ is the halo bias \citep{2001MNRAS.323....1S}. We then obtain the number of neighboring halos within radius $d$ and above $M^*$
\begin{equation}
N=\int_{<d} 4\pi r^2dr\int_{>M^*} [1+\xi_h(r,z,m_1,M_h)]\frac{dn}{dM_{\rm h}}dM_{\rm h};
\end{equation}
$d$ is then determined by solving the above equation for $N=1$.

The bubble model uses the cumulative ionizing photons number per collapsed atom, $\zeta$, to calculate ${\mathcal P}_b$, but this number is not explicitly appearing in our algorithm. However, we find that although the reionization history depends on $\zeta$, if we plot the ${\mathcal P}_b$ as a function of $\zeta f_{\rm coll}$ ($\sim Q_{\rm HII}$), where $f_{\rm coll}$ is the collapse fraction, we actually see limited variations in a large $\zeta$ range (see different curves in Fig. \ref{fig_Pb}). Therefore, in this paper we adopt the following approximation:  when calculating the ${\mathcal P}_b$ we fix $\zeta = 10$ and take the ${\mathcal P}_b$ value at the time at which $\zeta f_{\rm coll}=Q_{\rm HII}$. 
The above algorithm does not take into account the dependence of ${\mathcal P}_b$ on the halo formation redshift, $z_f$. Halos formed earlier are more biased, therefore they might have higher probability to be located in the ionized bubbles. This improvement will be deferred to future work. 

\end{document}